\documentclass[aps,prl,twocolumn,nofootinbib,superscriptaddress]{revtex4-1}
\usepackage{graphicx}
\usepackage{subfigure}
\usepackage{bbm}
\usepackage{amsmath}
\usepackage{amssymb}
\usepackage{amsfonts}
\usepackage{mathrsfs}
\usepackage{theorem}
\usepackage{bm}
\usepackage{url}
\usepackage[T1]{fontenc}
\usepackage{csquotes}
\MakeOuterQuote{"}

\usepackage{algorithm}
\usepackage{algorithmicx}
\usepackage{algpseudocode}

\usepackage{dcolumn}
\usepackage{color}
\usepackage[colorlinks,citecolor=blue]{hyperref}

\newcommand{\Rnum}[1]{\mathrm{\uppercase\expandafter{\romannumeral #1\relax}}}

\newcommand{\be}{\begin{equation}}
\newcommand{\ee}{\end{equation}}
\newcommand{\ba}{\begin{align}}
\newcommand{\ea}{\end{align}}

\def\<{\langle}  
\def\>{\rangle}  

\def\eqref#1{\textup{(\ref{#1})}}  
\newcommand{\eref}[1]{Eq.~\textup{(\ref{#1})}}

\newcommand{\esref}[1]{Eqs.~\textup{(\ref{#1})}}

\newcommand{\cref}[1]{Conjecture~\ref{#1}}
\newcommand{\Cref}[1]{Conjecture~\ref{#1}}

\begin{document}

\title{Efficient Experimental Verification of Quantum Gates with Local Operations}

\author{Rui-Qi Zhang}
\thanks{These authors contributed equally to this work.}
\affiliation{Key Laboratory of Quantum Information,University of Science and Technology of China, CAS, Hefei 230026, P. R. China}
\affiliation{Synergetic Innovation Center of Quantum Information and Quantum Physics, University of Science and Technology of China, Hefei 230026, P. R. China}
\author{Zhibo Hou}
\thanks{These authors contributed equally to this work.}
\affiliation{Key Laboratory of Quantum Information,University of Science and Technology of China, CAS, Hefei 230026, P. R. China}
\affiliation{Synergetic Innovation Center of Quantum Information and Quantum Physics, University of Science and Technology of China, Hefei 230026, P. R. China}
\author{Jun-Feng Tang}
\affiliation{Key Laboratory of Quantum Information,University of Science and Technology of China, CAS, Hefei 230026, P. R. China}
\affiliation{Synergetic Innovation Center of Quantum Information and Quantum Physics, University of Science and Technology of China, Hefei 230026, P. R. China}
\author{Jiangwei Shang}
\email{jiangwei.shang@bit.edu.cn}
\affiliation{Key Laboratory of Advanced Optoelectronic Quantum Architecture and Measurement of Ministry of Education, School of Physics, Beijing Institute of Technology, Beijing 100081, China}
\author{Huangjun~Zhu}
\email{zhuhuangjun@fudan.edu.cn}
\affiliation{State Key Laboratory of Surface Physics and Department of Physics, Fudan University, Shanghai 200433, China}
\affiliation{Institute for Nanoelectronic Devices and Quantum Computing, Fudan University, Shanghai 200433, China}
\affiliation{Center for Field Theory and Particle Physics, Fudan University, Shanghai 200433, China}
\author{Guo-Yong Xiang}
\email{gyxiang@ustc.edu.cn}
\affiliation{Key Laboratory of Quantum Information,University of Science and Technology of China, CAS, Hefei 230026, P. R. China}
\affiliation{Synergetic Innovation Center of Quantum Information and Quantum Physics, University of Science and Technology of China, Hefei 230026, P. R. China}
\author{Chuan-Feng Li}
\affiliation{Key Laboratory of Quantum Information,University of Science and Technology of China, CAS, Hefei 230026, P. R. China}
\affiliation{Synergetic Innovation Center of Quantum Information and Quantum Physics, University of Science and Technology of China, Hefei 230026, P. R. China}
\author{Guang-Can Guo}
\affiliation{Key Laboratory of Quantum Information,University of Science and Technology of China, CAS, Hefei 230026, P. R. China}
\affiliation{Synergetic Innovation Center of Quantum Information and Quantum Physics, University of Science and Technology of China, Hefei 230026, P. R. China}

\begin{abstract}
Verifying the correct functioning of quantum gates
is a crucial step towards reliable quantum information processing,
but it becomes an overwhelming challenge as the system size grows due to the dimensionality curse.
Recent theoretical breakthroughs show that it is possible to verify various important quantum gates
with the optimal sample complexity of $O(1/\epsilon)$ 
using local operations only, where $\epsilon$ is the estimation precision. In this work, we propose a variant of quantum gate verification (QGV) which is
robust to practical gate imperfections, and experimentally realize
efficient QGV on a two-qubit controlled-not gate and a three-qubit Toffoli gate using
only local state preparations and measurements. The experimental results show that,
by using only 1600 and 2600 measurements on average, we can verify with
$95$\% confidence level that the implemented controlled-not gate and Toffoli gate
have fidelities at least $99$\% and $97$\%, respectively. Demonstrating the superior low sample complexity and experimental feasibility of QGV, our work promises a solution to the dimensionality curse in verifying large quantum devices in the quantum era. 
\end{abstract}

\date{\today}

\maketitle

\textit{Introduction.---}%
Quantum computers can perform computational tasks much more efficiently \cite{Quantum_linear_eq, grover}
and even exponentially faster than their classical counterparts
\cite{shor, boson_sample, jiuzhang}.
Before harnessing the power of a quantum computer, a crucial
step is to
verify the correct functioning of its building blocks, i.e., the
quantum gates.
Traditional quantum process tomography (QPT) \cite{QPT1, QPT2} can provide the complete information of a quantum gate and  is a feasible solution for small systems. However, QPT is
not scalable as its complexity grows exponentially with
the size of the quantum system, and so far has been applied to quantum gates
acting on no more than three qubits \cite{QPT_EXP_2Q_1, QPT_EXP_2Q_2, QPT_EXP_3Q_1}. This exponential resource cost cannot be circumvented in general
even if one can take advantage of  the sparsity 
of the underlying structures
\cite{QST_CS1, QST_CS2, QST_MP} or heuristic algorithms \cite{SGQT_thoery, SGQT_exp}.

The key observation towards efficient verification of a quantum gate is that
the complete information of a quantum gate is usually not necessary in many tasks.
Quite often the fidelity of a quantum gate is enough to
characterize its quality.
Fidelity estimation based on unitary 2-designs and the twirling
protocol \cite{twirl1, 2_design_exp}
can estimate the fidelity of a Clifford gate with
size-independent sample complexity of $O(1/\epsilon^2)$,
where $\epsilon$ is the estimation precision.
Direct fidelity estimation and Monte Carlo sampling
\cite{MC1, MC2, MC_exp} can achieve a similar  sample complexity for Clifford and other well-conditioned gates
even if one can only prepare product states and perform Pauli measurements. 
Randomized benchmarking (RB) \cite{RB_interleave1, RB_interleave2, RB_individual, RB_exp_nonclifford}
can  certify Clifford gates and some special non-Clifford gates with a similar sample complexity,
and possesses the additional advantage of robustness against state-preparation and measurement errors.

Despite the progresses mentioned above,
most approaches in the literature  have disadvantages, which limit their applicability. Notably, most approaches are limited to  a few types of quantum gates (say Clifford gates) \cite{twirl1, 2_design_exp, MC1, MC2, MC_exp, RB_interleave1}. In addition, they have a suboptimal
scaling behavior in the precision $\epsilon$.
Moreover, many approaches, including
twirling protocols and RB,
require entangling  operations \cite{twirl1, 2_design_exp, RB_interleave1, RB_interleave2, RB_individual, RB_exp_nonclifford}, that is, preparing entangled states or performing entangling measurements.

Recently, an alternative approach called quantum gate 
verification (QGV) or quantum process verification (QPV) 
\cite{JWSHANG1,HGZHU1,ZengZL20} has been developed to tackle these problems. It is  inspired by probabilistic verification protocols which have found fruitful applications in certifying  quantum states
\cite{HayaMT06, QSV2, QSV4, QSV5, QSV6} and entanglement
\cite{ent_detec1, ent_detec2}.  
With this approach, a variety of quantum
gates can be verified efficiently with the optimal sample complexity of
$O(1/\epsilon)$ using only local state preparations and measurements. Nevertheless,
the current formulation of QGV can reach a valid conclusion only when the gate to be verified passes all the tests, which may prevent QGV from
obtaining a valid conclusion when a realistic quantum gate
with acceptable infidelity is considered.

In this Letter, we propose a variant of QGV which is
tolerant to gate imperfections, while keeping its
efficiency. With this robust proposal,
we experimentally apply QGV to a two-qubit controlled-not ({\sc cnot}) gate realized in a photonic system. By using 20 experimental settings and 1600 samples on
average we can verify
that the {\sc cnot} gate has at least $99$\% fidelity with a $95$\% confidence level.
We then  apply QGV to a three-qubit Toffoli gate to illustrate the scalability and
superiority of QGV. By using  32 measurement settings  and 2600 samples on average we can  verify that the fidelity of the
Toffoli gate is at least $97$\% with a $95$\% confidence
level. By contrast, the standard QPT would require at least 4096
measurement settings and over a million measurements in total to characterize the Toffoli gate.
Our experiments demonstrate that efficient verification of quantum gates can be
achieved with only local state preparations and measurements.

\textit{Theoretical framework.---}%
Consider a quantum device that is expected to implement a  target unitary
transformation $\mathcal{U}$, but actually realizes $N$ unknown quantum channels
$\Lambda_1, ..., \Lambda_N$, which are assumed to be identical and independent, over the $N$
runs. In practice, these channels might deviate from $\mathcal{U}$.
Let $1-\epsilon_A$ be the average gate fidelity of the channels with respect to $\mathcal{U}$. Our goal is to verify, with some confidence level $1-\delta$ (significance level $\delta$),
that the average gate infidelity of the channels is not larger than a given
threshold $\epsilon$, i.e.,
\begin{equation}\label{eq:criterion}
	\epsilon_{A} \le \epsilon\,,\quad {\rm with\ confidence\ level}\ 1 - \delta\,.	
\end{equation}

\begin{figure}[t]
	\center{\includegraphics[scale=0.28]{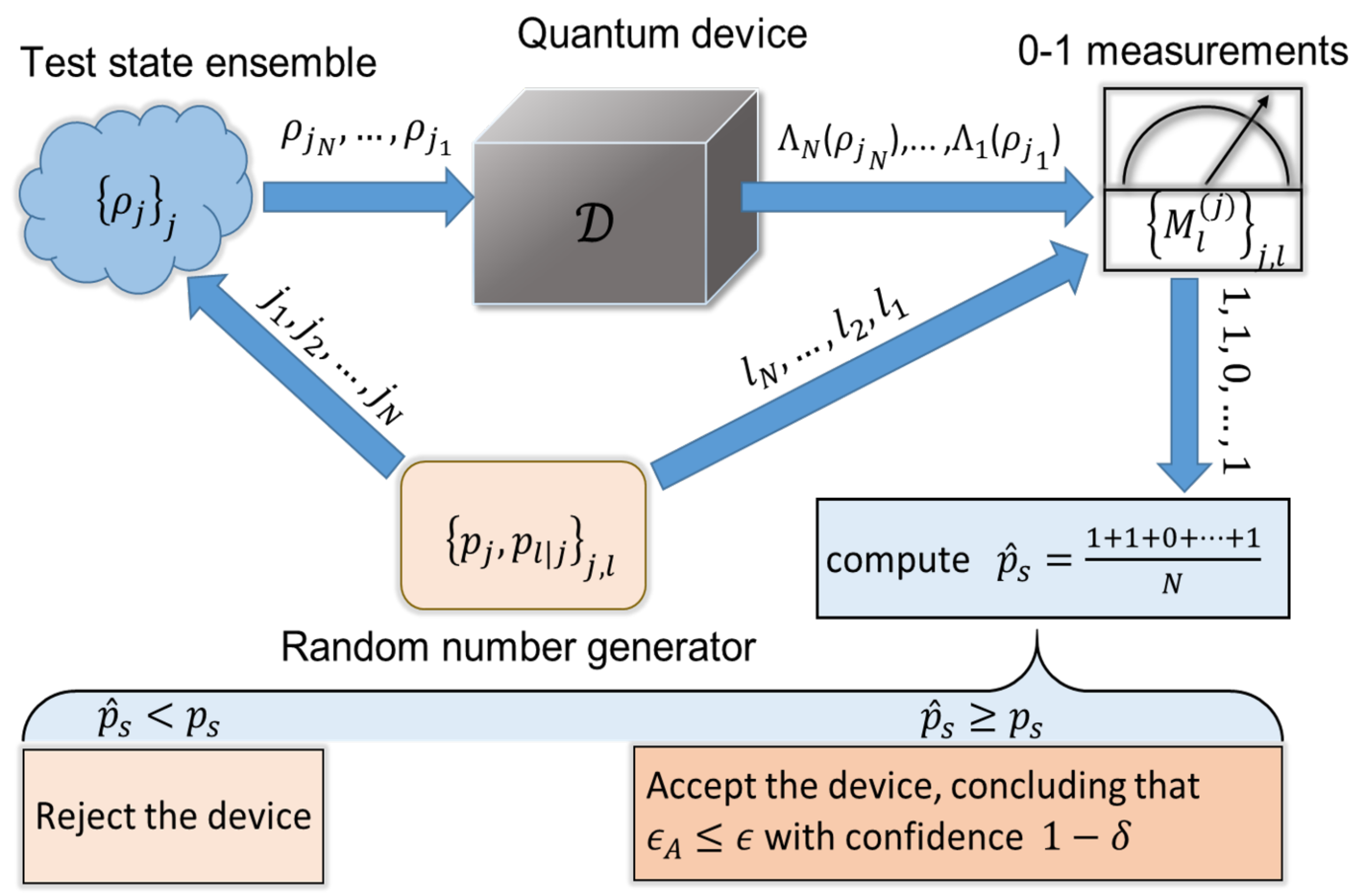}}
	\caption{\label{fig:scheme}
	Procedure for verifying the quantum device $\mathcal{D}$. In each run, the
	random number generator generates random numbers $j$ and $l$ according to the probabilities
	$p_j$ and $p_{l|j}$ (conditioned on $j$), respectively. Then state $\rho_j$ is drawn from the set of test states and sent to $\mathcal{D}$; next, the measurement
	module implements a two-outcome measurement $\{M_l^{(j)}, \openone-M_l^{(j)}\}$
	on the output state $\Lambda(\rho_j)$.
	By repeating the above procedure $N$ times, the verifier can reach a
	conclusion on the quality of $\mathcal{D}$ based on the passing frequency
	$\hat{p}_s$ over the $N$ tests.
	}	
\end{figure}

The verification procedure, illustrated in Fig.~\ref{fig:scheme},
can be described as follows \cite{HGZHU1}. In  the $i$th run,
the verifier first randomly chooses a pure state $\rho_j = |\psi_j\rangle\langle\psi_j|$ with probability
$p_j$ from a set of test states $\{\rho_j \}_j$ and subjects it to the device. Then
 the verifier performs a two-outcome  measurement
$\{M_l^{(j)}, \openone - M_l^{(j)}\}$,  which is called a test, on the output state $\Lambda_i (\rho_j)$ with outcome 1 for
passing and 0 for failure.
Here the test operator $M_l^{(j)}$ needs to satisfy the condition ${\rm Tr}[M_l^{(j)}\mathcal{U}(\rho_j)] = 1$ and is chosen randomly
with the conditional probability $p_{l|j}$ from a test set
$\{M_l^{(j)} \}_l$ that depends on $\mathcal{U}(\rho_j)$.
The verifier records the test results of the $N$ runs and compares the passing rate $\hat{p}_s$ with a given threshold $p_s$,
based on which the device is accepted or  rejected.

The performance of the above verification procedure is mainly determined by the \textit{process verification operator} defined as \cite{HGZHU1} \begin{equation}\label{eq:verification_op}
\Theta := d\sum_{j}{p_j \,\mathcal{U}^{-1}\biggl(\sum_{l}{p_{l|j} M_{l}^{(j)}}\biggr)\otimes\rho_j^{*}}\,.
\end{equation}
For a perfect device, the acceptance probability is unity. 
If the quantum gate realized has (average gate) infidelity $\epsilon$, by contrast,  the  acceptance probability is upper bounded by $[p_A (\Theta,\epsilon)]^N$, where $p_A (\Theta,\epsilon)$ is defined as the maximal passing
probability for quantum gates with infidelity $\epsilon_{A} \geq  \epsilon$ given the verification operator $\Theta$ \cite{HGZHU1}. If we set $p_s=1$, then
the minimal number of tests required to
verify the quantum gate with infidelity $\epsilon$ and confidence level $1-\delta$ reads
\begin{equation}\label{eq:generalScale}
N(\epsilon, \delta, \Theta) =
\biggl\lceil \frac{\ln \delta}{\ln p_A (\Theta, \epsilon)} \biggr\rceil.
\end{equation}
This number is minimized when
the test states $\rho_j$ form a 2-design \cite{2design1,2design2} and the test operator for each test state $\rho_j$ is chosen to be the projector
$\mathcal{U}(\rho_j)$ onto the target output state, in which case $p_A (\Theta,\epsilon)=1-\epsilon$,
and \eref{eq:generalScale} reduces to \cite{HGZHU1}
\begin{equation}\label{eq:optscale}
N^{\rm opt}(\epsilon, \delta) =
\biggl\lceil \frac{\ln \delta}{\ln (1 - \epsilon)} \biggr\rceil
\mathop{\approx} \limits^{\epsilon \to 0}
\frac{\ln \delta^{-1}}{\epsilon}\,.
\end{equation}

In general, to realize the optimal verification protocol mentioned above would require entangling operations, which are often inaccessible. 
Fortunately, for many important quantum  gates, nearly optimal performance can be achieved 
using local state preparations and local projective measurements only \cite{HGZHU1,ZengZL20,JWSHANG1}. 
For simplicity, in this work we focus on verification protocols that are balanced, which means the set of test states satisfies the condition $\sum_j p_j \rho_j=\openone/d$, where $d$ is the dimension of the underlying Hilbert space.
Denote by $\nu:=\nu(\Theta)$ the spectral gap of $\Theta$ (between the largest and the second largest eigenvalues), then we have
\begin{equation}
    N^{\rm local}(\epsilon, \delta, \Theta)
    \le \biggl\lceil \frac{\ln\delta}{\ln(1 - \nu\epsilon)} \biggr\rceil
    \le \biggl\lceil \frac{\ln\delta^{-1}}{\nu\epsilon} \biggl\rceil.
\end{equation}

In practice, quantum gates are never perfect. Even if they satisfy the condition
${\epsilon_{A} \le \epsilon}$, a few failure events might happen with a
non-negligible probability among the $N$ tests. In this case, setting $p_s=1$ for the threshold would reject a properly
functioning device with certain probability.
To remedy this problem and construct a robust verification protocol, we need to consider the situation with ${p_s<1}$. To be concrete, if
the passing frequency $\hat{p}_s$ over the $N$ tests is larger than $p_A(\Theta,\epsilon)$,
then the confidence level $1-\delta(\hat{p}_s)$ that the device satisfies $\epsilon_{A} \le \epsilon$
is lower bounded by
\begin{equation}\label{eq:chernoff}
	1-\delta(\hat{p}_s) \ge 1 - e^{-D(\hat{p}_s \| p_A (\Theta,\epsilon))N}\,,
\end{equation}
where $D(x \| y)=x\ln(\frac{x}{y})+(1-x)\ln(\frac{1-x}{1-y})$ is the Kullback-Leibler
 divergence.
On the other hand, given the confidence level $1 - \delta$, we can derive from \eref{eq:chernoff} an upper
bound for the infidelity $\epsilon_{A}$,
\begin{equation}\label{eq:infidelity}
	\epsilon_{A} \le \frac{d}{d+1}\frac{1-D^{(-1)}(\hat{p}_s, \ln{\delta^{-1}/N)}}{\nu(\Theta)}\,,
\end{equation}
where $D^{(-1)}(\hat{p}_s, y)$ is the inverse function of $y=D(\hat{p}_s \| x)$  with domain ${0 \le x < \hat{p}_s}$ (for a fixed  $\hat{p}_s$). The
detailed derivations of \esref{eq:chernoff} and \eqref{eq:infidelity} are
relegated  to Sec.~S1 in the Supplemental Material \cite{supp}.

\begin{figure*}[htbp]
	\center{\includegraphics[scale=0.18]{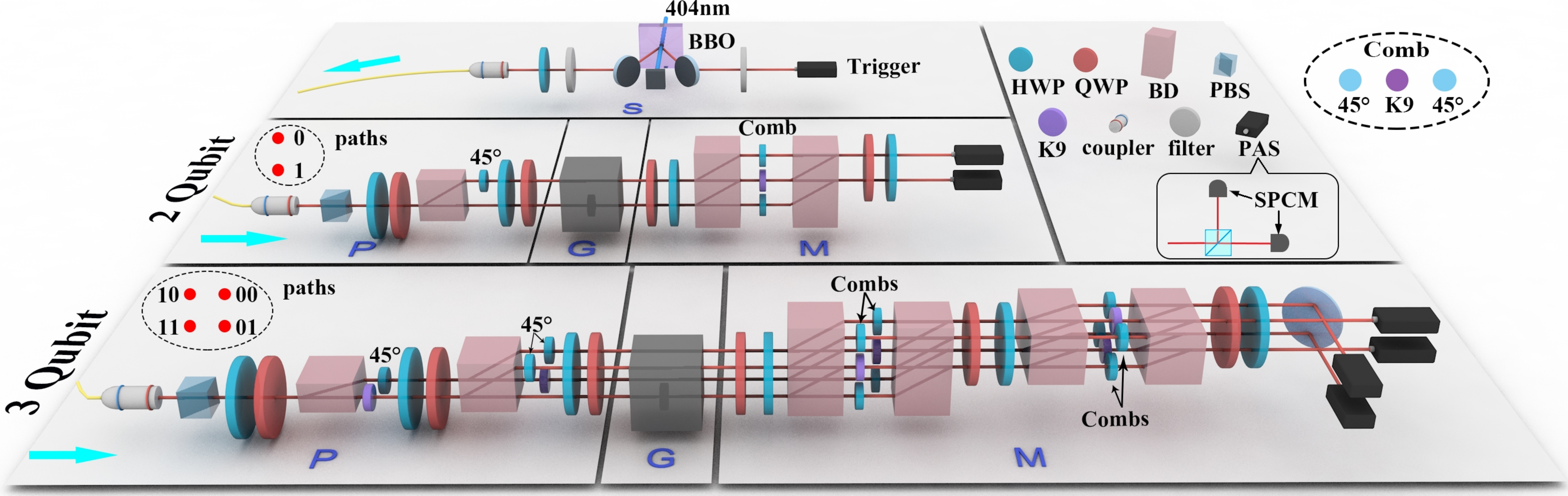}}
	\caption{\label{fig:setup}
	Experimental setup. The heralded single-photon source
	(labeled by S) is realized by spontaneous parametric
	down-conversion in a type-I $\beta$-barium-borate (BBO) crystal.
	The figure shows two independent setups employed for  implementing the verification protocols for the two-qubit {\sc cnot} gate and three-qubit Toffoli gate,  respectively. Each setup consists of three modules: a state-preparation
	module (labeled by P), a quantum-gate module (labeled by G), and
	a measurement module (labeled by M).  The inset in the component
	panel (upper right) shows the details of the polarization analyzing
	system (PAS). Each PAS consists of one polarizing
	beam-splitter (PBS) and two single-photon
	counting modules (SPCMs) and can  measure the photons in the
	$\{|H\rangle, |V\rangle\}$ polarization basis. HWP: half-wave
	plate; QWP: quarter-wave plate; BD: beam displacer;
    K9: K9 plate.
	}
\end{figure*}
\textit{Experimental setup.---}%
The experimental setups for verifying two-qubit
and three-qubit quantum gates are shown in
Fig.~\ref{fig:setup}. Both of them
consist of three modules: a state-preparation
module, a quantum-gate module, and a measurement module.
Here we use the  path and polarization degrees of freedom (DoFs)
of the heralded photon to encode the test state employed in
the verification.  The two-qubit system consists of a path DoF with up and down  modes   and  a polarization
DoF with horizontal (H) and vertical (V)
polarizations;  by contrast, the three-qubit system consists of a path DoF with left-right modes and up-down modes and a polarization DoF.

The heralded single-photon source
shown in Fig.~\ref{fig:setup} is used by both setups. 
An ultraviolet laser with central wavelength of $404$nm is used to pump
a type-I phase-matched $\beta$-barium-borate (BBO) crystal
to generate a photon pair in the product (polarization)
state via spontaneous parametric down-conversion \cite{photoSource}.
One photon is measured as a trigger to herald the generation of its
twin photon, which is then transmitted to the state-preparation module.

The state-preparation module in the two-qubit (three-qubit) setup is designed to
 prepare arbitrary two-qubit (three-qubit)  product states by
virtue of photonic quantum walks. Here the coin operators required are realized by combinations of half-wave plates (HWPs) and quarter-wave plates (QWPs); see Sec.~S2 of the Supplemental Material \cite{supp}.
The K9 plates in the state-preparation
module in the three-qubit setup are used to
compensate for the path-length difference among the interference arms.

The quantum-gate module implements the quantum gate to be verified, which can be seen
as a black box that is expected to perform the target unitary transformation on
the input quantum states. The measurement module in the two-qubit (three-qubit) setup is
designed to realize arbitrary local projective measurements on two-qubit (three-qubit)
systems by using photonic quantum walks.
The QWP-HWP pairs inside the measurement
module control the measurement settings for individual qubits;
see Sec.~S2 of the Supplemental Material \cite{supp}.
In addition, the K9 plates
are used to compensate for the path-length difference among the interference arms. Finally, the heralded
photon is collected by two polarization analyzing systems (PASs)
in the two-qubit setup and four PASs in the three-qubit setup, where the PASs
measure the polarization of the input photon in the
$\{|H\rangle, |V\rangle\}$ basis.

\textit{Results.---}%
To demonstrate the efficiency and  scalability of QGV, we performed QGV  on a two-qubit {\sc cnot} gate and a three-qubit Toffoli gate. The {\sc cnot} gate (Toffoli gate)
is implemented by inserting a HWP with its optical axis aligned at
$45^\circ$ to the horizontal direction on path $1$ ($11$) in the
two-qubit (three-qubit)  setup.
The sets of test states and measurement settings employed for verifying the
{\sc cnot} gate and Toffoli gate are detailed
in Sec.~S3 of the Supplemental Material \cite{supp}.

The performance of QGV is characterized by the scalings of the significance level 
$\delta$ and infidelity $\epsilon_{A}$ with respect to the number of tests $N$.
The values of $\delta$ and $\epsilon_{A}$ after each test
can be determined  from
the test results by virtue of  \esref{eq:chernoff} and \eqref{eq:infidelity}.
Since the results of a single run of QGV suffer
from statistical fluctuations, which would prevent us from reliably evaluating
 the performance, we repeat the verification procedure 50 times under the same conditions
(e.g., the set of test states and the number of tests in total).
The average values of $\delta$ and $\epsilon_{A}$ are calculated
by substituting $\hat{p}_s$ in \esref{eq:chernoff} and \eqref{eq:infidelity} with
$\sum_{i=1}^{50}{\hat{p}_{s}^{(i)}}/50$ for each value of $N$, where $\hat{p}_{s}^{(i)}$ is the
passing rate of the $i$th run among the first $N$ tests. We also use \esref{eq:chernoff} and
\eqref{eq:infidelity} to fit the average results by fixing the value of $\hat{p}_s$
to be the average passing rate over
the $50$ runs of QGV among all the tests used.

\begin{figure}[t]
	\center{\includegraphics[scale=0.30]{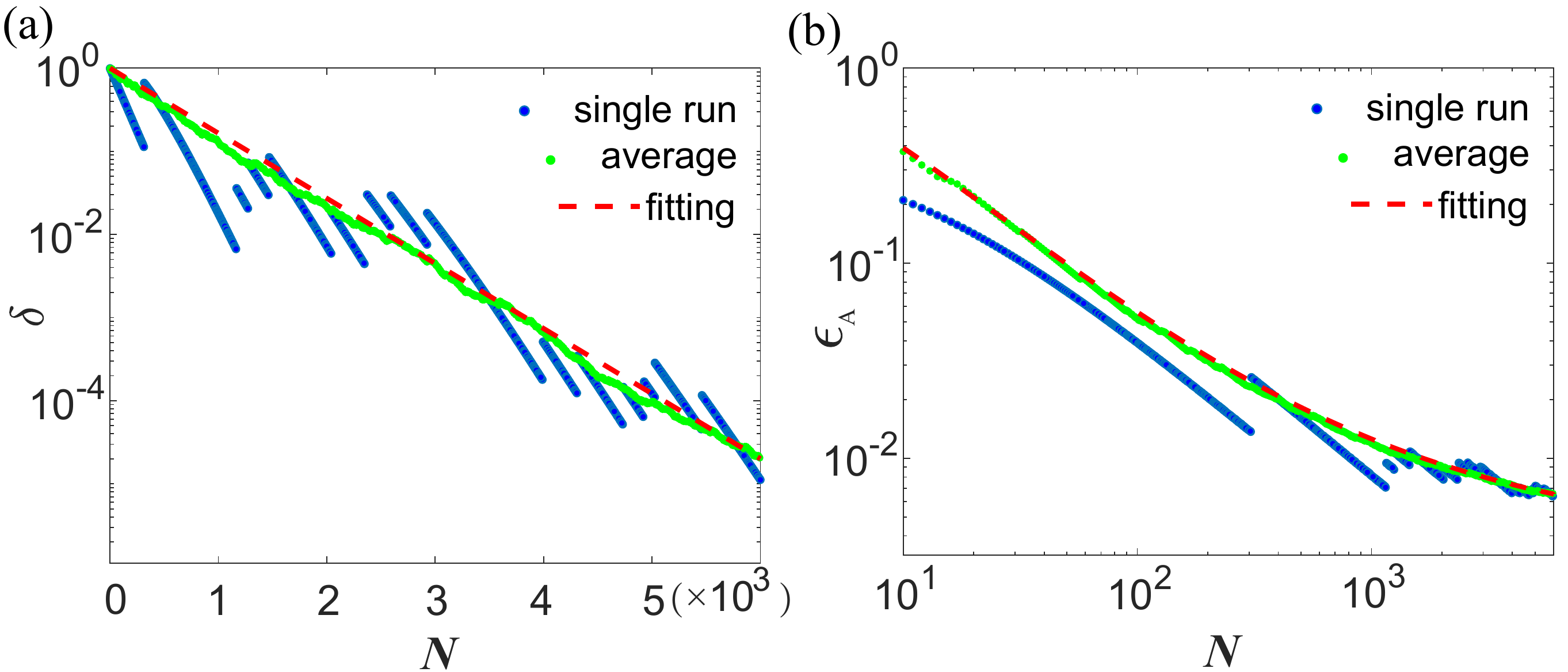}}
	\caption{\label{fig:QGV_results_for_CNOT}
	Experimental results on  the verification of the {\sc cnot} gate. The blue dots represent
	the results of a single run of QGV. The green dots
	represent the average results of $50$ runs of QGV. The red
	dotted line is the fitting line for the average results.
	(a) When $\epsilon$ is set to $0.01$, $\delta$ is log plotted versus
	$N$. (b) When $\delta$ is set to $0.05$, $\epsilon_{A}$ is
	log-log plotted versus $N$. Within the first $200$ tests,  the scaling of
	$\epsilon_A$ averaged over 50 runs with respect to $N$  
	is fitted to be $N^{-0.857}$ by linear regression.
    }
\end{figure}

The experimental results on the verification of the {\sc cnot} gate are shown in
Fig.~\ref{fig:QGV_results_for_CNOT}, where $20$ different measurement settings and $6000$ tests in total
 (see Sec.~S3 of the Supplemental
Material \cite{supp} for details) are used in each run of QGV.
In Fig.~\ref{fig:QGV_results_for_CNOT}(a), where $\epsilon$ is set to be $0.01$,
$\delta$ rapidly drops below $0.05$ within $1600$ tests for both the single-run
and average results, which means that the {\sc cnot} gate is verified efficiently
with high confidence level. Alternatively, we can set the confidence level
$1-\delta$ to be $0.95$ and calculate $\epsilon_{A}$.
Fig.~\ref{fig:QGV_results_for_CNOT}(b) shows that $\epsilon_{A}$ descends below $0.01$ after $1600$ tests
for both the single-run and average results, which is consistent with
Fig.~\ref{fig:QGV_results_for_CNOT}(a). The scaling of the average infidelity
$\epsilon_{A}$ with respect to $N$ can be described by the power law $N^{-0.857}$ within the first $200$ tests, which is quite close to the optimal scaling of $N^{-1}$ in
\eref{eq:optscale}. After $200$ tests the descending speed of $\epsilon_{A}$ gradually slows down
as it gets closer to the actual infidelity, and eventually converges
to $0.0045$ after $10000$ tests (see Sec.~S4 of the Supplemental
Material \cite{supp}).
In both Fig.~3(a) and (b), the single-run results  break up into discrete
short segments due to the occasional failures caused by the deviation of the actual gate
from the ideal target gate.

We then perform QPT on the {\sc cnot} gate and find that the actual
average gate fidelity is $99.7\%$, which is consistent with the QGV result.
To perform QPT on the {\sc cnot} gate, we
employ 36 product Pauli eigenstates as the test states and 9 measurement settings based on
Pauli measurements for each output state.
The experimental details are relegated to Sec.~S5 of the Supplemental
Material \cite{supp}.
Here the total number of experimental settings is  $324$, and the total number of measurements is over 6 million, which are substantially more than that required in QGV (the number of measurements in QPT can be reduced, but the conclusion does not change). These facts  clearly reflect the advantage of QGV over QPT.

\begin{figure}[t]
	\center{\includegraphics[scale=0.30]{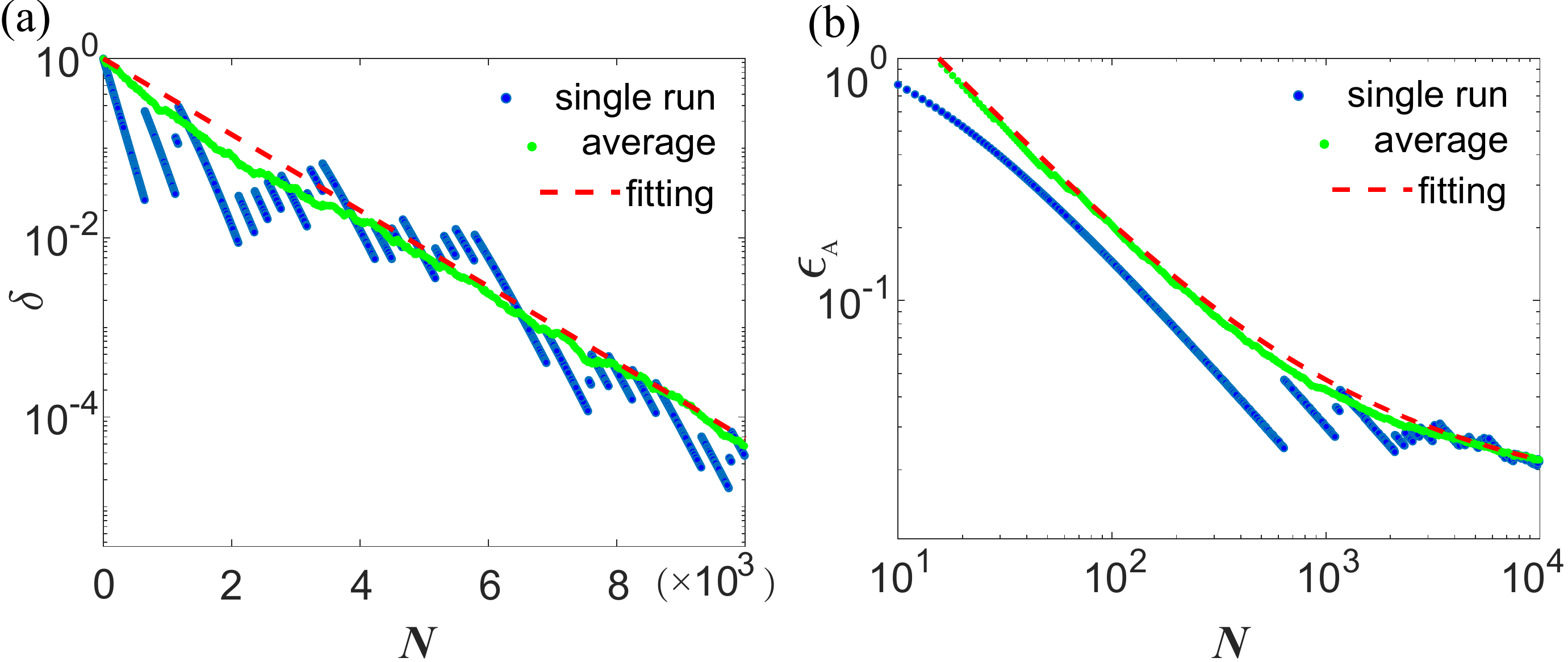}}
	\caption{\label{fig:QGV_results_for_Toffoli_gate}
Experimental  results on the verification of  the Toffoli gate. The meanings of the data points
	are similar to those in Fig.~\ref{fig:QGV_results_for_CNOT}.
	(a) When $\epsilon$ is set to $0.03$, $\delta$ is log plotted versus $N$.
	(b) When $\delta$ is set to $0.05$, $\epsilon_{A}$ is log-log plotted
	versus $N$.  Within the first $200$ tests, the scaling of	$\epsilon_A$ averaged over 50 runs with respect to $N$ 	is fitted to be $N^{-0.840}$ by linear regression.
	}
\end{figure}

To demonstrate the scalability of QGV, next we consider the verification of the three-qubit Toffoli gate.
In this case, $32$ different experimental  settings and $10000$ tests in total  are employed  in each run of QGV (see Sec.~S3 of
the Supplemental Material \cite{supp} for details).
The verification results are shown in Fig.~\ref{fig:QGV_results_for_Toffoli_gate}, which are analogous to the counterparts shown in  Fig.~\ref{fig:QGV_results_for_CNOT}.
To  verify the Toffoli
gate within infidelity  $0.03$ and confidence level
 $95$\%, only 2600 tests are required.
In Fig.~\ref{fig:QGV_results_for_Toffoli_gate}(b), $\epsilon_{A}$
exhibits $N^{-0.840}$ scaling with respect to $N$ within the first $200$ tests, which is
also close to the optimal scaling of $N^{-1}$.
The infidelity estimator $\epsilon_{A}$ eventually converges to
$0.0148$ after $40000$ tests; see Sec.~S4 of the Supplemental Material \cite{supp}.
In both plots in Fig.~\ref{fig:QGV_results_for_Toffoli_gate}, the single-run results break up more frequently
than their  counterparts in Fig.~\ref{fig:QGV_results_for_CNOT}, due to the larger deviation of the
actual Toffoli gate from the ideal Toffoli gate. Incidentally,
to  perform QPT on the Toffoli gate would require  ${8^4=4096}$ experimental settings and millions of measurements in total, which are quite prohibitive and  much more resource consuming than QGV.

\textit{Summary.---}By virtue of photonic systems, we experimentally realized efficient verification of a {\sc cnot} gate and a Toffoli gate
with local state preparations and measurements. The experimental results
clearly show that  the verification protocols can achieve nearly optimal performance without relying on entangling operations, and are substantially more efficient than QPT. Moreover, they are scalable and robust to the imperfections of the actual quantum gates.
Notably, only 2600 tests are required to verify the Toffoli gate with fidelity 97\% and confidence level 95\%.
 Our work demonstrates that QGV is a powerful tool for the verification of quantum gates and quantum devices, and may play a key role in the development of quantum technologies.

\section*{ACKNOWLEDGMENTS}
The work at the University of Science and Technology of China is supported by the National Natural Science Foundation of China (Grants Nos. 61905234, 11974335, 11574291, and 11774334), the Key Research Program of Frontier Sciences, CAS (Grant No. QYZDYSSW-SLH003) and the Fundamental Research Funds for the Central Universities (Grant No. WK2470000035). JS acknowledges support by the Beijing Institute of Technology Research Fund Program for Young Scholars and the National Natural Science Foundation of China (Grant No. 11805010).
The work at Fudan University is supported by  the National Natural Science Foundation of China (Grant No.~11875110) and  Shanghai Municipal Science and Technology Major Project (Grant No.~2019SHZDZX01). 

\bibliography{cites}

\begin{thebibliography}{38}%
\makeatletter
\providecommand \@ifxundefined [1]{%
 \@ifx{#1\undefined}
}%
\providecommand \@ifnum [1]{%
 \ifnum #1\expandafter \@firstoftwo
 \else \expandafter \@secondoftwo
 \fi
}%
\providecommand \@ifx [1]{%
 \ifx #1\expandafter \@firstoftwo
 \else \expandafter \@secondoftwo
 \fi
}%
\providecommand \natexlab [1]{#1}%
\providecommand \enquote  [1]{``#1''}%
\providecommand \bibnamefont  [1]{#1}%
\providecommand \bibfnamefont [1]{#1}%
\providecommand \citenamefont [1]{#1}%
\providecommand \href@noop [0]{\@secondoftwo}%
\providecommand \href [0]{\begingroup \@sanitize@url \@href}%
\providecommand \@href[1]{\@@startlink{#1}\@@href}%
\providecommand \@@href[1]{\endgroup#1\@@endlink}%
\providecommand \@sanitize@url [0]{\catcode `\\12\catcode `\$12\catcode
  `\&12\catcode `\#12\catcode `\^12\catcode `\_12\catcode `\%12\relax}%
\providecommand \@@startlink[1]{}%
\providecommand \@@endlink[0]{}%
\providecommand \url  [0]{\begingroup\@sanitize@url \@url }%
\providecommand \@url [1]{\endgroup\@href {#1}{\urlprefix }}%
\providecommand \urlprefix  [0]{URL }%
\providecommand \Eprint [0]{\href }%
\providecommand \doibase [0]{http://dx.doi.org/}%
\providecommand \selectlanguage [0]{\@gobble}%
\providecommand \bibinfo  [0]{\@secondoftwo}%
\providecommand \bibfield  [0]{\@secondoftwo}%
\providecommand \translation [1]{[#1]}%
\providecommand \BibitemOpen [0]{}%
\providecommand \bibitemStop [0]{}%
\providecommand \bibitemNoStop [0]{.\EOS\space}%
\providecommand \EOS [0]{\spacefactor3000\relax}%
\providecommand \BibitemShut  [1]{\csname bibitem#1\endcsname}%
\let\auto@bib@innerbib\@empty
\bibitem [{\citenamefont {Harrow}\ \emph {et~al.}(2009)\citenamefont {Harrow},
  \citenamefont {Hassidim},\ and\ \citenamefont {Lloyd}}]{Quantum_linear_eq}%
  \BibitemOpen
  \bibfield  {author} {\bibinfo {author} {\bibfnamefont {A.~W.}\ \bibnamefont
  {Harrow}}, \bibinfo {author} {\bibfnamefont {A.}~\bibnamefont {Hassidim}}, \
  and\ \bibinfo {author} {\bibfnamefont {S.}~\bibnamefont {Lloyd}},\ }\href
  {\doibase 10.1103/PhysRevLett.103.150502} {\bibfield  {journal} {\bibinfo
  {journal} {Phys. Rev. Lett.}\ }\textbf {\bibinfo {volume} {103}},\ \bibinfo
  {pages} {150502} (\bibinfo {year} {2009})}\BibitemShut {NoStop}%
\bibitem [{\citenamefont {Grover}(1996)}]{grover}%
  \BibitemOpen
  \bibfield  {author} {\bibinfo {author} {\bibfnamefont {L.~K.}\ \bibnamefont
  {Grover}},\ }in\ \href {\doibase 10.1145/237814.237866} {\emph {\bibinfo
  {booktitle} {Proceedings of the Twenty-Eighth Annual ACM Symposium on Theory
  of Computing}}},\ \bibinfo {series and number} {STOC '96}\ (\bibinfo
  {publisher} {Association for Computing Machinery},\ \bibinfo {address} {New
  York, NY, USA},\ \bibinfo {year} {1996})\ p.\ \bibinfo {pages}
  {212–219}\BibitemShut {NoStop}%
\bibitem [{\citenamefont {Shor}(1994)}]{shor}%
  \BibitemOpen
  \bibfield  {author} {\bibinfo {author} {\bibfnamefont {P.}~\bibnamefont
  {Shor}},\ }in\ \href {\doibase 10.1109/SFCS.1994.365700} {\emph {\bibinfo
  {booktitle} {Proceedings of the 35th Annual Symposium on Foundations of
  Computer Science}}}\ (\bibinfo {year} {1994})\ pp.\ \bibinfo {pages}
  {124--134}\BibitemShut {NoStop}%
\bibitem [{\citenamefont {Aaronson}\ and\ \citenamefont
  {Arkhipov}(2011)}]{boson_sample}%
  \BibitemOpen
  \bibfield  {author} {\bibinfo {author} {\bibfnamefont {S.}~\bibnamefont
  {Aaronson}}\ and\ \bibinfo {author} {\bibfnamefont {A.}~\bibnamefont
  {Arkhipov}},\ }in\ \href {\doibase 10.1145/1993636.1993682} {\emph {\bibinfo
  {booktitle} {Proceedings of the Forty-Third Annual ACM Symposium on Theory of
  Computing}}},\ \bibinfo {series and number} {STOC '11}\ (\bibinfo
  {publisher} {Association for Computing Machinery},\ \bibinfo {address} {New
  York, NY, USA},\ \bibinfo {year} {2011})\ p.\ \bibinfo {pages}
  {333–342}\BibitemShut {NoStop}%
\bibitem [{\citenamefont {Zhong}\ \emph {et~al.}(2020)\citenamefont {Zhong},
  \citenamefont {Wang}, \citenamefont {Deng}, \citenamefont {Chen},
  \citenamefont {Peng}, \citenamefont {Luo}, \citenamefont {Qin}, \citenamefont
  {Wu}, \citenamefont {Ding}, \citenamefont {Hu}, \citenamefont {Hu},
  \citenamefont {Yang}, \citenamefont {Zhang}, \citenamefont {Li},
  \citenamefont {Li}, \citenamefont {Jiang}, \citenamefont {Gan}, \citenamefont
  {Yang}, \citenamefont {You}, \citenamefont {Wang}, \citenamefont {Li},
  \citenamefont {Liu}, \citenamefont {Lu},\ and\ \citenamefont
  {Pan}}]{jiuzhang}%
  \BibitemOpen
  \bibfield  {author} {\bibinfo {author} {\bibfnamefont {H.-S.}\ \bibnamefont
  {Zhong}}, \bibinfo {author} {\bibfnamefont {H.}~\bibnamefont {Wang}},
  \bibinfo {author} {\bibfnamefont {Y.-H.}\ \bibnamefont {Deng}}, \bibinfo
  {author} {\bibfnamefont {M.-C.}\ \bibnamefont {Chen}}, \bibinfo {author}
  {\bibfnamefont {L.-C.}\ \bibnamefont {Peng}}, \bibinfo {author}
  {\bibfnamefont {Y.-H.}\ \bibnamefont {Luo}}, \bibinfo {author} {\bibfnamefont
  {J.}~\bibnamefont {Qin}}, \bibinfo {author} {\bibfnamefont {D.}~\bibnamefont
  {Wu}}, \bibinfo {author} {\bibfnamefont {X.}~\bibnamefont {Ding}}, \bibinfo
  {author} {\bibfnamefont {Y.}~\bibnamefont {Hu}}, \bibinfo {author}
  {\bibfnamefont {P.}~\bibnamefont {Hu}}, \bibinfo {author} {\bibfnamefont
  {X.-Y.}\ \bibnamefont {Yang}}, \bibinfo {author} {\bibfnamefont {W.-J.}\
  \bibnamefont {Zhang}}, \bibinfo {author} {\bibfnamefont {H.}~\bibnamefont
  {Li}}, \bibinfo {author} {\bibfnamefont {Y.}~\bibnamefont {Li}}, \bibinfo
  {author} {\bibfnamefont {X.}~\bibnamefont {Jiang}}, \bibinfo {author}
  {\bibfnamefont {L.}~\bibnamefont {Gan}}, \bibinfo {author} {\bibfnamefont
  {G.}~\bibnamefont {Yang}}, \bibinfo {author} {\bibfnamefont {L.}~\bibnamefont
  {You}}, \bibinfo {author} {\bibfnamefont {Z.}~\bibnamefont {Wang}}, \bibinfo
  {author} {\bibfnamefont {L.}~\bibnamefont {Li}}, \bibinfo {author}
  {\bibfnamefont {N.-L.}\ \bibnamefont {Liu}}, \bibinfo {author} {\bibfnamefont
  {C.-Y.}\ \bibnamefont {Lu}}, \ and\ \bibinfo {author} {\bibfnamefont {J.-W.}\
  \bibnamefont {Pan}},\ }\href {\doibase 10.1126/science.abe8770} {\bibfield
  {journal} {\bibinfo  {journal} {Science}\ }\textbf {\bibinfo {volume}
  {370}},\ \bibinfo {pages} {1460} (\bibinfo {year} {2020})}\BibitemShut
  {NoStop}%
\bibitem [{\citenamefont {Chuang}\ and\ \citenamefont {Nielsen}(1997)}]{QPT1}%
  \BibitemOpen
  \bibfield  {author} {\bibinfo {author} {\bibfnamefont {I.~L.}\ \bibnamefont
  {Chuang}}\ and\ \bibinfo {author} {\bibfnamefont {M.~A.}\ \bibnamefont
  {Nielsen}},\ }\href {\doibase 10.1080/09500349708231894} {\bibfield
  {journal} {\bibinfo  {journal} {J. Mod. Opt.}\ }\textbf {\bibinfo {volume}
  {44}},\ \bibinfo {pages} {2455} (\bibinfo {year} {1997})}\BibitemShut
  {NoStop}%
\bibitem [{\citenamefont {Poyatos}\ \emph {et~al.}(1997)\citenamefont
  {Poyatos}, \citenamefont {Cirac},\ and\ \citenamefont {Zoller}}]{QPT2}%
  \BibitemOpen
  \bibfield  {author} {\bibinfo {author} {\bibfnamefont {J.~F.}\ \bibnamefont
  {Poyatos}}, \bibinfo {author} {\bibfnamefont {J.~I.}\ \bibnamefont {Cirac}},
  \ and\ \bibinfo {author} {\bibfnamefont {P.}~\bibnamefont {Zoller}},\ }\href
  {\doibase 10.1103/PhysRevLett.78.390} {\bibfield  {journal} {\bibinfo
  {journal} {Phys. Rev. Lett.}\ }\textbf {\bibinfo {volume} {78}},\ \bibinfo
  {pages} {390} (\bibinfo {year} {1997})}\BibitemShut {NoStop}%
\bibitem [{\citenamefont {Riebe}\ \emph {et~al.}(2006)\citenamefont {Riebe},
  \citenamefont {Kim}, \citenamefont {Schindler}, \citenamefont {Monz},
  \citenamefont {Schmidt}, \citenamefont {K\"orber}, \citenamefont {H\"ansel},
  \citenamefont {H\"affner}, \citenamefont {Roos},\ and\ \citenamefont
  {Blatt}}]{QPT_EXP_2Q_1}%
  \BibitemOpen
  \bibfield  {author} {\bibinfo {author} {\bibfnamefont {M.}~\bibnamefont
  {Riebe}}, \bibinfo {author} {\bibfnamefont {K.}~\bibnamefont {Kim}}, \bibinfo
  {author} {\bibfnamefont {P.}~\bibnamefont {Schindler}}, \bibinfo {author}
  {\bibfnamefont {T.}~\bibnamefont {Monz}}, \bibinfo {author} {\bibfnamefont
  {P.~O.}\ \bibnamefont {Schmidt}}, \bibinfo {author} {\bibfnamefont {T.~K.}\
  \bibnamefont {K\"orber}}, \bibinfo {author} {\bibfnamefont {W.}~\bibnamefont
  {H\"ansel}}, \bibinfo {author} {\bibfnamefont {H.}~\bibnamefont {H\"affner}},
  \bibinfo {author} {\bibfnamefont {C.~F.}\ \bibnamefont {Roos}}, \ and\
  \bibinfo {author} {\bibfnamefont {R.}~\bibnamefont {Blatt}},\ }\href
  {\doibase 10.1103/PhysRevLett.97.220407} {\bibfield  {journal} {\bibinfo
  {journal} {Phys. Rev. Lett.}\ }\textbf {\bibinfo {volume} {97}},\ \bibinfo
  {pages} {220407} (\bibinfo {year} {2006})}\BibitemShut {NoStop}%
\bibitem [{\citenamefont {Bialczak}\ \emph {et~al.}(2010)\citenamefont
  {Bialczak}, \citenamefont {Ansmann}, \citenamefont {Hofheinz}, \citenamefont
  {Lucero}, \citenamefont {Neeley}, \citenamefont {O’Connell}, \citenamefont
  {Sank}, \citenamefont {Wang}, \citenamefont {Wenner}, \citenamefont
  {Steffen}, \citenamefont {Cleland},\ and\ \citenamefont
  {Martinis}}]{QPT_EXP_2Q_2}%
  \BibitemOpen
  \bibfield  {author} {\bibinfo {author} {\bibfnamefont {R.~C.}\ \bibnamefont
  {Bialczak}}, \bibinfo {author} {\bibfnamefont {M.}~\bibnamefont {Ansmann}},
  \bibinfo {author} {\bibfnamefont {M.}~\bibnamefont {Hofheinz}}, \bibinfo
  {author} {\bibfnamefont {E.}~\bibnamefont {Lucero}}, \bibinfo {author}
  {\bibfnamefont {M.}~\bibnamefont {Neeley}}, \bibinfo {author} {\bibfnamefont
  {A.~D.}\ \bibnamefont {O’Connell}}, \bibinfo {author} {\bibfnamefont
  {D.}~\bibnamefont {Sank}}, \bibinfo {author} {\bibfnamefont {H.}~\bibnamefont
  {Wang}}, \bibinfo {author} {\bibfnamefont {J.}~\bibnamefont {Wenner}},
  \bibinfo {author} {\bibfnamefont {M.}~\bibnamefont {Steffen}}, \bibinfo
  {author} {\bibfnamefont {A.~N.}\ \bibnamefont {Cleland}}, \ and\ \bibinfo
  {author} {\bibfnamefont {J.~M.}\ \bibnamefont {Martinis}},\ }\href {\doibase
  10.1038/nphys1639} {\bibfield  {journal} {\bibinfo  {journal} {Nat. Phys.}\
  }\textbf {\bibinfo {volume} {6}},\ \bibinfo {pages} {409} (\bibinfo {year}
  {2010})}\BibitemShut {NoStop}%
\bibitem [{\citenamefont {Weinstein}\ \emph {et~al.}(2004)\citenamefont
  {Weinstein}, \citenamefont {Havel}, \citenamefont {Emerson}, \citenamefont
  {Boulant}, \citenamefont {Saraceno}, \citenamefont {Lloyd},\ and\
  \citenamefont {Cory}}]{QPT_EXP_3Q_1}%
  \BibitemOpen
  \bibfield  {author} {\bibinfo {author} {\bibfnamefont {Y.~S.}\ \bibnamefont
  {Weinstein}}, \bibinfo {author} {\bibfnamefont {T.~F.}\ \bibnamefont
  {Havel}}, \bibinfo {author} {\bibfnamefont {J.}~\bibnamefont {Emerson}},
  \bibinfo {author} {\bibfnamefont {N.}~\bibnamefont {Boulant}}, \bibinfo
  {author} {\bibfnamefont {M.}~\bibnamefont {Saraceno}}, \bibinfo {author}
  {\bibfnamefont {S.}~\bibnamefont {Lloyd}}, \ and\ \bibinfo {author}
  {\bibfnamefont {D.~G.}\ \bibnamefont {Cory}},\ }\href {\doibase
  10.1063/1.1785151} {\bibfield  {journal} {\bibinfo  {journal} {J. Chem.
  Phys.}\ }\textbf {\bibinfo {volume} {121}},\ \bibinfo {pages} {6117}
  (\bibinfo {year} {2004})}\BibitemShut {NoStop}%
\bibitem [{\citenamefont {Gross}\ \emph {et~al.}(2010)\citenamefont {Gross},
  \citenamefont {Liu}, \citenamefont {Flammia}, \citenamefont {Becker},\ and\
  \citenamefont {Eisert}}]{QST_CS1}%
  \BibitemOpen
  \bibfield  {author} {\bibinfo {author} {\bibfnamefont {D.}~\bibnamefont
  {Gross}}, \bibinfo {author} {\bibfnamefont {Y.-K.}\ \bibnamefont {Liu}},
  \bibinfo {author} {\bibfnamefont {S.~T.}\ \bibnamefont {Flammia}}, \bibinfo
  {author} {\bibfnamefont {S.}~\bibnamefont {Becker}}, \ and\ \bibinfo {author}
  {\bibfnamefont {J.}~\bibnamefont {Eisert}},\ }\href {\doibase
  10.1103/PhysRevLett.105.150401} {\bibfield  {journal} {\bibinfo  {journal}
  {Phys. Rev. Lett.}\ }\textbf {\bibinfo {volume} {105}},\ \bibinfo {pages}
  {150401} (\bibinfo {year} {2010})}\BibitemShut {NoStop}%
\bibitem [{\citenamefont {Flammia}\ \emph {et~al.}(2012)\citenamefont
  {Flammia}, \citenamefont {Gross}, \citenamefont {Liu},\ and\ \citenamefont
  {Eisert}}]{QST_CS2}%
  \BibitemOpen
  \bibfield  {author} {\bibinfo {author} {\bibfnamefont {S.~T.}\ \bibnamefont
  {Flammia}}, \bibinfo {author} {\bibfnamefont {D.}~\bibnamefont {Gross}},
  \bibinfo {author} {\bibfnamefont {Y.-K.}\ \bibnamefont {Liu}}, \ and\
  \bibinfo {author} {\bibfnamefont {J.}~\bibnamefont {Eisert}},\ }\href
  {\doibase 10.1088/1367-2630/14/9/095022} {\bibfield  {journal} {\bibinfo
  {journal} {New J. Phys.}\ }\textbf {\bibinfo {volume} {14}},\ \bibinfo
  {pages} {095022} (\bibinfo {year} {2012})}\BibitemShut {NoStop}%
\bibitem [{\citenamefont {Cramer}\ \emph {et~al.}(2010)\citenamefont {Cramer},
  \citenamefont {Plenio}, \citenamefont {Flammia}, \citenamefont {Somma},
  \citenamefont {Gross}, \citenamefont {Bartlett}, \citenamefont
  {Landon-Cardinal}, \citenamefont {Poulin},\ and\ \citenamefont
  {Liu}}]{QST_MP}%
  \BibitemOpen
  \bibfield  {author} {\bibinfo {author} {\bibfnamefont {M.}~\bibnamefont
  {Cramer}}, \bibinfo {author} {\bibfnamefont {M.~B.}\ \bibnamefont {Plenio}},
  \bibinfo {author} {\bibfnamefont {S.~T.}\ \bibnamefont {Flammia}}, \bibinfo
  {author} {\bibfnamefont {R.}~\bibnamefont {Somma}}, \bibinfo {author}
  {\bibfnamefont {D.}~\bibnamefont {Gross}}, \bibinfo {author} {\bibfnamefont
  {S.~D.}\ \bibnamefont {Bartlett}}, \bibinfo {author} {\bibfnamefont
  {O.}~\bibnamefont {Landon-Cardinal}}, \bibinfo {author} {\bibfnamefont
  {D.}~\bibnamefont {Poulin}}, \ and\ \bibinfo {author} {\bibfnamefont {Y.-K.}\
  \bibnamefont {Liu}},\ }\href {\doibase 10.1038/ncomms1147} {\bibfield
  {journal} {\bibinfo  {journal} {Nat. Commun.}\ }\textbf {\bibinfo {volume}
  {1}},\ \bibinfo {pages} {149} (\bibinfo {year} {2010})}\BibitemShut {NoStop}%
\bibitem [{\citenamefont {Ferrie}(2014)}]{SGQT_thoery}%
  \BibitemOpen
  \bibfield  {author} {\bibinfo {author} {\bibfnamefont {C.}~\bibnamefont
  {Ferrie}},\ }\href {\doibase 10.1103/PhysRevLett.113.190404} {\bibfield
  {journal} {\bibinfo  {journal} {Phys. Rev. Lett.}\ }\textbf {\bibinfo
  {volume} {113}},\ \bibinfo {pages} {190404} (\bibinfo {year}
  {2014})}\BibitemShut {NoStop}%
\bibitem [{\citenamefont {Hou}\ \emph {et~al.}(2020)\citenamefont {Hou},
  \citenamefont {Tang}, \citenamefont {Ferrie}, \citenamefont {Xiang},
  \citenamefont {Li},\ and\ \citenamefont {Guo}}]{SGQT_exp}%
  \BibitemOpen
  \bibfield  {author} {\bibinfo {author} {\bibfnamefont {Z.}~\bibnamefont
  {Hou}}, \bibinfo {author} {\bibfnamefont {J.-F.}\ \bibnamefont {Tang}},
  \bibinfo {author} {\bibfnamefont {C.}~\bibnamefont {Ferrie}}, \bibinfo
  {author} {\bibfnamefont {G.-Y.}\ \bibnamefont {Xiang}}, \bibinfo {author}
  {\bibfnamefont {C.-F.}\ \bibnamefont {Li}}, \ and\ \bibinfo {author}
  {\bibfnamefont {G.-C.}\ \bibnamefont {Guo}},\ }\href {\doibase
  10.1103/PhysRevA.101.022317} {\bibfield  {journal} {\bibinfo  {journal}
  {Phys. Rev. A}\ }\textbf {\bibinfo {volume} {101}},\ \bibinfo {pages}
  {022317} (\bibinfo {year} {2020})}\BibitemShut {NoStop}%
\bibitem [{\citenamefont {Dankert}\ \emph {et~al.}(2009)\citenamefont
  {Dankert}, \citenamefont {Cleve}, \citenamefont {Emerson},\ and\
  \citenamefont {Livine}}]{twirl1}%
  \BibitemOpen
  \bibfield  {author} {\bibinfo {author} {\bibfnamefont {C.}~\bibnamefont
  {Dankert}}, \bibinfo {author} {\bibfnamefont {R.}~\bibnamefont {Cleve}},
  \bibinfo {author} {\bibfnamefont {J.}~\bibnamefont {Emerson}}, \ and\
  \bibinfo {author} {\bibfnamefont {E.}~\bibnamefont {Livine}},\ }\href
  {\doibase 10.1103/PhysRevA.80.012304} {\bibfield  {journal} {\bibinfo
  {journal} {Phys. Rev. A}\ }\textbf {\bibinfo {volume} {80}},\ \bibinfo
  {pages} {012304} (\bibinfo {year} {2009})}\BibitemShut {NoStop}%
\bibitem [{\citenamefont {Lu}\ \emph {et~al.}(2015)\citenamefont {Lu},
  \citenamefont {Li}, \citenamefont {Trottier}, \citenamefont {Li},
  \citenamefont {Brodutch}, \citenamefont {Krismanich}, \citenamefont
  {Ghavami}, \citenamefont {Dmitrienko}, \citenamefont {Long}, \citenamefont
  {Baugh},\ and\ \citenamefont {Laflamme}}]{2_design_exp}%
  \BibitemOpen
  \bibfield  {author} {\bibinfo {author} {\bibfnamefont {D.}~\bibnamefont
  {Lu}}, \bibinfo {author} {\bibfnamefont {H.}~\bibnamefont {Li}}, \bibinfo
  {author} {\bibfnamefont {D.-A.}\ \bibnamefont {Trottier}}, \bibinfo {author}
  {\bibfnamefont {J.}~\bibnamefont {Li}}, \bibinfo {author} {\bibfnamefont
  {A.}~\bibnamefont {Brodutch}}, \bibinfo {author} {\bibfnamefont {A.~P.}\
  \bibnamefont {Krismanich}}, \bibinfo {author} {\bibfnamefont
  {A.}~\bibnamefont {Ghavami}}, \bibinfo {author} {\bibfnamefont {G.~I.}\
  \bibnamefont {Dmitrienko}}, \bibinfo {author} {\bibfnamefont
  {G.}~\bibnamefont {Long}}, \bibinfo {author} {\bibfnamefont {J.}~\bibnamefont
  {Baugh}}, \ and\ \bibinfo {author} {\bibfnamefont {R.}~\bibnamefont
  {Laflamme}},\ }\href {\doibase 10.1103/PhysRevLett.114.140505} {\bibfield
  {journal} {\bibinfo  {journal} {Phys. Rev. Lett.}\ }\textbf {\bibinfo
  {volume} {114}},\ \bibinfo {pages} {140505} (\bibinfo {year}
  {2015})}\BibitemShut {NoStop}%
\bibitem [{\citenamefont {Flammia}\ and\ \citenamefont {Liu}(2011)}]{MC1}%
  \BibitemOpen
  \bibfield  {author} {\bibinfo {author} {\bibfnamefont {S.~T.}\ \bibnamefont
  {Flammia}}\ and\ \bibinfo {author} {\bibfnamefont {Y.-K.}\ \bibnamefont
  {Liu}},\ }\href {\doibase 10.1103/PhysRevLett.106.230501} {\bibfield
  {journal} {\bibinfo  {journal} {Phys. Rev. Lett.}\ }\textbf {\bibinfo
  {volume} {106}},\ \bibinfo {pages} {230501} (\bibinfo {year}
  {2011})}\BibitemShut {NoStop}%
\bibitem [{\citenamefont {da~Silva}\ \emph {et~al.}(2011)\citenamefont
  {da~Silva}, \citenamefont {Landon-Cardinal},\ and\ \citenamefont
  {Poulin}}]{MC2}%
  \BibitemOpen
  \bibfield  {author} {\bibinfo {author} {\bibfnamefont {M.~P.}\ \bibnamefont
  {da~Silva}}, \bibinfo {author} {\bibfnamefont {O.}~\bibnamefont
  {Landon-Cardinal}}, \ and\ \bibinfo {author} {\bibfnamefont {D.}~\bibnamefont
  {Poulin}},\ }\href {\doibase 10.1103/PhysRevLett.107.210404} {\bibfield
  {journal} {\bibinfo  {journal} {Phys. Rev. Lett.}\ }\textbf {\bibinfo
  {volume} {107}},\ \bibinfo {pages} {210404} (\bibinfo {year}
  {2011})}\BibitemShut {NoStop}%
\bibitem [{\citenamefont {Steffen}\ \emph {et~al.}(2012)\citenamefont
  {Steffen}, \citenamefont {da~Silva}, \citenamefont {Fedorov}, \citenamefont
  {Baur},\ and\ \citenamefont {Wallraff}}]{MC_exp}%
  \BibitemOpen
  \bibfield  {author} {\bibinfo {author} {\bibfnamefont {L.}~\bibnamefont
  {Steffen}}, \bibinfo {author} {\bibfnamefont {M.~P.}\ \bibnamefont
  {da~Silva}}, \bibinfo {author} {\bibfnamefont {A.}~\bibnamefont {Fedorov}},
  \bibinfo {author} {\bibfnamefont {M.}~\bibnamefont {Baur}}, \ and\ \bibinfo
  {author} {\bibfnamefont {A.}~\bibnamefont {Wallraff}},\ }\href {\doibase
  10.1103/PhysRevLett.108.260506} {\bibfield  {journal} {\bibinfo  {journal}
  {Phys. Rev. Lett.}\ }\textbf {\bibinfo {volume} {108}},\ \bibinfo {pages}
  {260506} (\bibinfo {year} {2012})}\BibitemShut {NoStop}%
\bibitem [{\citenamefont {Magesan}\ \emph {et~al.}(2012)\citenamefont
  {Magesan}, \citenamefont {Gambetta}, \citenamefont {Johnson}, \citenamefont
  {Ryan}, \citenamefont {Chow}, \citenamefont {Merkel}, \citenamefont
  {da~Silva}, \citenamefont {Keefe}, \citenamefont {Rothwell}, \citenamefont
  {Ohki}, \citenamefont {Ketchen},\ and\ \citenamefont
  {Steffen}}]{RB_interleave1}%
  \BibitemOpen
  \bibfield  {author} {\bibinfo {author} {\bibfnamefont {E.}~\bibnamefont
  {Magesan}}, \bibinfo {author} {\bibfnamefont {J.~M.}\ \bibnamefont
  {Gambetta}}, \bibinfo {author} {\bibfnamefont {B.~R.}\ \bibnamefont
  {Johnson}}, \bibinfo {author} {\bibfnamefont {C.~A.}\ \bibnamefont {Ryan}},
  \bibinfo {author} {\bibfnamefont {J.~M.}\ \bibnamefont {Chow}}, \bibinfo
  {author} {\bibfnamefont {S.~T.}\ \bibnamefont {Merkel}}, \bibinfo {author}
  {\bibfnamefont {M.~P.}\ \bibnamefont {da~Silva}}, \bibinfo {author}
  {\bibfnamefont {G.~A.}\ \bibnamefont {Keefe}}, \bibinfo {author}
  {\bibfnamefont {M.~B.}\ \bibnamefont {Rothwell}}, \bibinfo {author}
  {\bibfnamefont {T.~A.}\ \bibnamefont {Ohki}}, \bibinfo {author}
  {\bibfnamefont {M.~B.}\ \bibnamefont {Ketchen}}, \ and\ \bibinfo {author}
  {\bibfnamefont {M.}~\bibnamefont {Steffen}},\ }\href {\doibase
  10.1103/PhysRevLett.109.080505} {\bibfield  {journal} {\bibinfo  {journal}
  {Phys. Rev. Lett.}\ }\textbf {\bibinfo {volume} {109}},\ \bibinfo {pages}
  {080505} (\bibinfo {year} {2012})}\BibitemShut {NoStop}%
\bibitem [{\citenamefont {Harper}\ and\ \citenamefont
  {Flammia}(2017)}]{RB_interleave2}%
  \BibitemOpen
  \bibfield  {author} {\bibinfo {author} {\bibfnamefont {R.}~\bibnamefont
  {Harper}}\ and\ \bibinfo {author} {\bibfnamefont {S.~T.}\ \bibnamefont
  {Flammia}},\ }\href {\doibase 10.1088/2058-9565/aa5f8d} {\bibfield  {journal}
  {\bibinfo  {journal} {Quantum Sci. Technol.}\ }\textbf {\bibinfo {volume}
  {2}},\ \bibinfo {pages} {015008} (\bibinfo {year} {2017})}\BibitemShut
  {NoStop}%
\bibitem [{\citenamefont {Onorati}\ \emph {et~al.}(2019)\citenamefont
  {Onorati}, \citenamefont {Werner},\ and\ \citenamefont
  {Eisert}}]{RB_individual}%
  \BibitemOpen
  \bibfield  {author} {\bibinfo {author} {\bibfnamefont {E.}~\bibnamefont
  {Onorati}}, \bibinfo {author} {\bibfnamefont {A.~H.}\ \bibnamefont {Werner}},
  \ and\ \bibinfo {author} {\bibfnamefont {J.}~\bibnamefont {Eisert}},\ }\href
  {\doibase 10.1103/PhysRevLett.123.060501} {\bibfield  {journal} {\bibinfo
  {journal} {Phys. Rev. Lett.}\ }\textbf {\bibinfo {volume} {123}},\ \bibinfo
  {pages} {060501} (\bibinfo {year} {2019})}\BibitemShut {NoStop}%
\bibitem [{\citenamefont {Garion}\ \emph {et~al.}(2021)\citenamefont {Garion},
  \citenamefont {Kanazawa}, \citenamefont {Landa}, \citenamefont {McKay},
  \citenamefont {Sheldon}, \citenamefont {Cross},\ and\ \citenamefont
  {Wood}}]{RB_exp_nonclifford}%
  \BibitemOpen
  \bibfield  {author} {\bibinfo {author} {\bibfnamefont {S.}~\bibnamefont
  {Garion}}, \bibinfo {author} {\bibfnamefont {N.}~\bibnamefont {Kanazawa}},
  \bibinfo {author} {\bibfnamefont {H.}~\bibnamefont {Landa}}, \bibinfo
  {author} {\bibfnamefont {D.~C.}\ \bibnamefont {McKay}}, \bibinfo {author}
  {\bibfnamefont {S.}~\bibnamefont {Sheldon}}, \bibinfo {author} {\bibfnamefont
  {A.~W.}\ \bibnamefont {Cross}}, \ and\ \bibinfo {author} {\bibfnamefont
  {C.~J.}\ \bibnamefont {Wood}},\ }\href {\doibase
  10.1103/PhysRevResearch.3.013204} {\bibfield  {journal} {\bibinfo  {journal}
  {Phys. Rev. Research}\ }\textbf {\bibinfo {volume} {3}},\ \bibinfo {pages}
  {013204} (\bibinfo {year} {2021})}\BibitemShut {NoStop}%
\bibitem [{\citenamefont {Liu}\ \emph {et~al.}(2020)\citenamefont {Liu},
  \citenamefont {Shang}, \citenamefont {Yu},\ and\ \citenamefont
  {Zhang}}]{JWSHANG1}%
  \BibitemOpen
  \bibfield  {author} {\bibinfo {author} {\bibfnamefont {Y.-C.}\ \bibnamefont
  {Liu}}, \bibinfo {author} {\bibfnamefont {J.}~\bibnamefont {Shang}}, \bibinfo
  {author} {\bibfnamefont {X.-D.}\ \bibnamefont {Yu}}, \ and\ \bibinfo {author}
  {\bibfnamefont {X.}~\bibnamefont {Zhang}},\ }\href {\doibase
  10.1103/PhysRevA.101.042315} {\bibfield  {journal} {\bibinfo  {journal}
  {Phys. Rev. A}\ }\textbf {\bibinfo {volume} {101}},\ \bibinfo {pages}
  {042315} (\bibinfo {year} {2020})}\BibitemShut {NoStop}%
\bibitem [{\citenamefont {Zhu}\ and\ \citenamefont {Zhang}(2020)}]{HGZHU1}%
  \BibitemOpen
  \bibfield  {author} {\bibinfo {author} {\bibfnamefont {H.}~\bibnamefont
  {Zhu}}\ and\ \bibinfo {author} {\bibfnamefont {H.}~\bibnamefont {Zhang}},\
  }\href {\doibase 10.1103/PhysRevA.101.042316} {\bibfield  {journal} {\bibinfo
   {journal} {Phys. Rev. A}\ }\textbf {\bibinfo {volume} {101}},\ \bibinfo
  {pages} {042316} (\bibinfo {year} {2020})}\BibitemShut {NoStop}%
\bibitem [{\citenamefont {Zeng}\ \emph {et~al.}(2020)\citenamefont {Zeng},
  \citenamefont {Zhou},\ and\ \citenamefont {Liu}}]{ZengZL20}%
  \BibitemOpen
  \bibfield  {author} {\bibinfo {author} {\bibfnamefont {P.}~\bibnamefont
  {Zeng}}, \bibinfo {author} {\bibfnamefont {Y.}~\bibnamefont {Zhou}}, \ and\
  \bibinfo {author} {\bibfnamefont {Z.}~\bibnamefont {Liu}},\ }\href {\doibase
  10.1103/PhysRevResearch.2.023306} {\bibfield  {journal} {\bibinfo  {journal}
  {Phys. Rev. Research}\ }\textbf {\bibinfo {volume} {2}},\ \bibinfo {pages}
  {023306} (\bibinfo {year} {2020})}\BibitemShut {NoStop}%
\bibitem [{\citenamefont {Hayashi}\ \emph {et~al.}(2006)\citenamefont
  {Hayashi}, \citenamefont {Matsumoto},\ and\ \citenamefont
  {Tsuda}}]{HayaMT06}%
  \BibitemOpen
  \bibfield  {author} {\bibinfo {author} {\bibfnamefont {M.}~\bibnamefont
  {Hayashi}}, \bibinfo {author} {\bibfnamefont {K.}~\bibnamefont {Matsumoto}},
  \ and\ \bibinfo {author} {\bibfnamefont {Y.}~\bibnamefont {Tsuda}},\
  }\href@noop {} {\bibfield  {journal} {\bibinfo  {journal} {J. Phys. A: Math.
  Gen.}\ }\textbf {\bibinfo {volume} {39}},\ \bibinfo {pages} {14427} (\bibinfo
  {year} {2006})}\BibitemShut {NoStop}%
\bibitem [{\citenamefont {Pallister}\ \emph {et~al.}(2018)\citenamefont
  {Pallister}, \citenamefont {Linden},\ and\ \citenamefont {Montanaro}}]{QSV2}%
  \BibitemOpen
  \bibfield  {author} {\bibinfo {author} {\bibfnamefont {S.}~\bibnamefont
  {Pallister}}, \bibinfo {author} {\bibfnamefont {N.}~\bibnamefont {Linden}}, \
  and\ \bibinfo {author} {\bibfnamefont {A.}~\bibnamefont {Montanaro}},\ }\href
  {\doibase 10.1103/PhysRevLett.120.170502} {\bibfield  {journal} {\bibinfo
  {journal} {Phys. Rev. Lett.}\ }\textbf {\bibinfo {volume} {120}},\ \bibinfo
  {pages} {170502} (\bibinfo {year} {2018})}\BibitemShut {NoStop}%
\bibitem [{\citenamefont {Zhang}\ \emph {et~al.}(2020)\citenamefont {Zhang},
  \citenamefont {Zhang}, \citenamefont {Chen}, \citenamefont {Peng},
  \citenamefont {Xu}, \citenamefont {Yin}, \citenamefont {Yu}, \citenamefont
  {Ye}, \citenamefont {Han}, \citenamefont {Xu}, \citenamefont {Chen},
  \citenamefont {Li},\ and\ \citenamefont {Guo}}]{QSV4}%
  \BibitemOpen
  \bibfield  {author} {\bibinfo {author} {\bibfnamefont {W.-H.}\ \bibnamefont
  {Zhang}}, \bibinfo {author} {\bibfnamefont {C.}~\bibnamefont {Zhang}},
  \bibinfo {author} {\bibfnamefont {Z.}~\bibnamefont {Chen}}, \bibinfo {author}
  {\bibfnamefont {X.-X.}\ \bibnamefont {Peng}}, \bibinfo {author}
  {\bibfnamefont {X.-Y.}\ \bibnamefont {Xu}}, \bibinfo {author} {\bibfnamefont
  {P.}~\bibnamefont {Yin}}, \bibinfo {author} {\bibfnamefont {S.}~\bibnamefont
  {Yu}}, \bibinfo {author} {\bibfnamefont {X.-J.}\ \bibnamefont {Ye}}, \bibinfo
  {author} {\bibfnamefont {Y.-J.}\ \bibnamefont {Han}}, \bibinfo {author}
  {\bibfnamefont {J.-S.}\ \bibnamefont {Xu}}, \bibinfo {author} {\bibfnamefont
  {G.}~\bibnamefont {Chen}}, \bibinfo {author} {\bibfnamefont {C.-F.}\
  \bibnamefont {Li}}, \ and\ \bibinfo {author} {\bibfnamefont {G.-C.}\
  \bibnamefont {Guo}},\ }\href {\doibase 10.1103/PhysRevLett.125.030506}
  {\bibfield  {journal} {\bibinfo  {journal} {Phys. Rev. Lett.}\ }\textbf
  {\bibinfo {volume} {125}},\ \bibinfo {pages} {030506} (\bibinfo {year}
  {2020})}\BibitemShut {NoStop}%
\bibitem [{\citenamefont {Zhu}\ and\ \citenamefont {Hayashi}(2019)}]{QSV5}%
  \BibitemOpen
  \bibfield  {author} {\bibinfo {author} {\bibfnamefont {H.}~\bibnamefont
  {Zhu}}\ and\ \bibinfo {author} {\bibfnamefont {M.}~\bibnamefont {Hayashi}},\
  }\href {\doibase 10.1103/PhysRevLett.123.260504} {\bibfield  {journal}
  {\bibinfo  {journal} {Phys. Rev. Lett.}\ }\textbf {\bibinfo {volume} {123}},\
  \bibinfo {pages} {260504} (\bibinfo {year} {2019})}\BibitemShut {NoStop}%
\bibitem [{\citenamefont {Liu}\ \emph {et~al.}(2021)\citenamefont {Liu},
  \citenamefont {Shang}, \citenamefont {Han},\ and\ \citenamefont
  {Zhang}}]{QSV6}%
  \BibitemOpen
  \bibfield  {author} {\bibinfo {author} {\bibfnamefont {Y.-C.}\ \bibnamefont
  {Liu}}, \bibinfo {author} {\bibfnamefont {J.}~\bibnamefont {Shang}}, \bibinfo
  {author} {\bibfnamefont {R.}~\bibnamefont {Han}}, \ and\ \bibinfo {author}
  {\bibfnamefont {X.}~\bibnamefont {Zhang}},\ }\href {\doibase
  10.1103/PhysRevLett.126.090504} {\bibfield  {journal} {\bibinfo  {journal}
  {Phys. Rev. Lett.}\ }\textbf {\bibinfo {volume} {126}},\ \bibinfo {pages}
  {090504} (\bibinfo {year} {2021})}\BibitemShut {NoStop}%
\bibitem [{\citenamefont {Dimić}\ and\ \citenamefont
  {Dakić}(2018)}]{ent_detec1}%
  \BibitemOpen
  \bibfield  {author} {\bibinfo {author} {\bibfnamefont {A.}~\bibnamefont
  {Dimić}}\ and\ \bibinfo {author} {\bibfnamefont {B.}~\bibnamefont
  {Dakić}},\ }\href {\doibase 10.1038/s41534-017-0055-x} {\bibfield  {journal}
  {\bibinfo  {journal} {npj Quantum Inf.}\ }\textbf {\bibinfo {volume} {4}},\
  \bibinfo {pages} {11} (\bibinfo {year} {2018})}\BibitemShut {NoStop}%
\bibitem [{\citenamefont {Saggio}\ \emph {et~al.}(2019)\citenamefont {Saggio},
  \citenamefont {Dimić}, \citenamefont {Greganti}, \citenamefont {Rozema},
  \citenamefont {Walther},\ and\ \citenamefont {Dakić}}]{ent_detec2}%
  \BibitemOpen
  \bibfield  {author} {\bibinfo {author} {\bibfnamefont {V.}~\bibnamefont
  {Saggio}}, \bibinfo {author} {\bibfnamefont {A.}~\bibnamefont {Dimić}},
  \bibinfo {author} {\bibfnamefont {C.}~\bibnamefont {Greganti}}, \bibinfo
  {author} {\bibfnamefont {L.~A.}\ \bibnamefont {Rozema}}, \bibinfo {author}
  {\bibfnamefont {P.}~\bibnamefont {Walther}}, \ and\ \bibinfo {author}
  {\bibfnamefont {B.}~\bibnamefont {Dakić}},\ }\href {\doibase
  10.1038/s41567-019-0550-4} {\bibfield  {journal} {\bibinfo  {journal} {Nat.
  Phys.}\ }\textbf {\bibinfo {volume} {15}},\ \bibinfo {pages} {935} (\bibinfo
  {year} {2019})}\BibitemShut {NoStop}%
\bibitem [{\citenamefont {Renes}\ \emph {et~al.}(2004)\citenamefont {Renes},
  \citenamefont {Blume-Kohout}, \citenamefont {Scott},\ and\ \citenamefont
  {Caves}}]{2design1}%
  \BibitemOpen
  \bibfield  {author} {\bibinfo {author} {\bibfnamefont {J.~M.}\ \bibnamefont
  {Renes}}, \bibinfo {author} {\bibfnamefont {R.}~\bibnamefont {Blume-Kohout}},
  \bibinfo {author} {\bibfnamefont {A.~J.}\ \bibnamefont {Scott}}, \ and\
  \bibinfo {author} {\bibfnamefont {C.~M.}\ \bibnamefont {Caves}},\ }\href
  {\doibase 10.1063/1.1737053} {\bibfield  {journal} {\bibinfo  {journal} {J.
  Math. Phys.}\ }\textbf {\bibinfo {volume} {45}},\ \bibinfo {pages} {2171}
  (\bibinfo {year} {2004})}\BibitemShut {NoStop}%
\bibitem [{\citenamefont {Roy}\ and\ \citenamefont {Scott}(2007)}]{2design2}%
  \BibitemOpen
  \bibfield  {author} {\bibinfo {author} {\bibfnamefont {A.}~\bibnamefont
  {Roy}}\ and\ \bibinfo {author} {\bibfnamefont {A.~J.}\ \bibnamefont
  {Scott}},\ }\href {\doibase 10.1063/1.2748617} {\bibfield  {journal}
  {\bibinfo  {journal} {J. Math. Phys.}\ }\textbf {\bibinfo {volume} {48}},\
  \bibinfo {pages} {072110} (\bibinfo {year} {2007})}\BibitemShut {NoStop}%
\bibitem [{sup()}]{supp}%
  \BibitemOpen
  \href@noop {} {}\bibinfo {note} {See Supplemental Material for the
  details.}\BibitemShut {Stop}%
\bibitem [{\citenamefont {Kwiat}\ \emph {et~al.}(1999)\citenamefont {Kwiat},
  \citenamefont {Waks}, \citenamefont {White}, \citenamefont {Appelbaum},\ and\
  \citenamefont {Eberhard}}]{photoSource}%
  \BibitemOpen
  \bibfield  {author} {\bibinfo {author} {\bibfnamefont {P.~G.}\ \bibnamefont
  {Kwiat}}, \bibinfo {author} {\bibfnamefont {E.}~\bibnamefont {Waks}},
  \bibinfo {author} {\bibfnamefont {A.~G.}\ \bibnamefont {White}}, \bibinfo
  {author} {\bibfnamefont {I.}~\bibnamefont {Appelbaum}}, \ and\ \bibinfo
  {author} {\bibfnamefont {P.~H.}\ \bibnamefont {Eberhard}},\ }\href {\doibase
  10.1103/PhysRevA.60.R773} {\bibfield  {journal} {\bibinfo  {journal} {Phys.
  Rev. A}\ }\textbf {\bibinfo {volume} {60}},\ \bibinfo {pages} {R773}
  (\bibinfo {year} {1999})}\BibitemShut {NoStop}%
\end{thebibliography}%


\begin{thebibliography}{7}%
\makeatletter
\providecommand \@ifxundefined [1]{%
 \@ifx{#1\undefined}
}%
\providecommand \@ifnum [1]{%
 \ifnum #1\expandafter \@firstoftwo
 \else \expandafter \@secondoftwo
 \fi
}%
\providecommand \@ifx [1]{%
 \ifx #1\expandafter \@firstoftwo
 \else \expandafter \@secondoftwo
 \fi
}%
\providecommand \natexlab [1]{#1}%
\providecommand \enquote  [1]{``#1''}%
\providecommand \bibnamefont  [1]{#1}%
\providecommand \bibfnamefont [1]{#1}%
\providecommand \citenamefont [1]{#1}%
\providecommand \href@noop [0]{\@secondoftwo}%
\providecommand \href [0]{\begingroup \@sanitize@url \@href}%
\providecommand \@href[1]{\@@startlink{#1}\@@href}%
\providecommand \@@href[1]{\endgroup#1\@@endlink}%
\providecommand \@sanitize@url [0]{\catcode `\\12\catcode `\$12\catcode
  `\&12\catcode `\#12\catcode `\^12\catcode `\_12\catcode `\%12\relax}%
\providecommand \@@startlink[1]{}%
\providecommand \@@endlink[0]{}%
\providecommand \url  [0]{\begingroup\@sanitize@url \@url }%
\providecommand \@url [1]{\endgroup\@href {#1}{\urlprefix }}%
\providecommand \urlprefix  [0]{URL }%
\providecommand \Eprint [0]{\href }%
\providecommand \doibase [0]{http://dx.doi.org/}%
\providecommand \selectlanguage [0]{\@gobble}%
\providecommand \bibinfo  [0]{\@secondoftwo}%
\providecommand \bibfield  [0]{\@secondoftwo}%
\providecommand \translation [1]{[#1]}%
\providecommand \BibitemOpen [0]{}%
\providecommand \bibitemStop [0]{}%
\providecommand \bibitemNoStop [0]{.\EOS\space}%
\providecommand \EOS [0]{\spacefactor3000\relax}%
\providecommand \BibitemShut  [1]{\csname bibitem#1\endcsname}%
\let\auto@bib@innerbib\@empty
\bibitem [{\citenamefont {Zhu}\ and\ \citenamefont {Zhang}(2020)}]{HGZHU1}%
  \BibitemOpen
  \bibfield  {author} {\bibinfo {author} {\bibfnamefont {H.}~\bibnamefont
  {Zhu}}\ and\ \bibinfo {author} {\bibfnamefont {H.}~\bibnamefont {Zhang}},\
  }\href {\doibase 10.1103/PhysRevA.101.042316} {\bibfield  {journal} {\bibinfo
   {journal} {Phys. Rev. A}\ }\textbf {\bibinfo {volume} {101}},\ \bibinfo
  {pages} {042316} (\bibinfo {year} {2020})}\BibitemShut {NoStop}%
\bibitem [{\citenamefont {Dimić}\ and\ \citenamefont
  {Dakić}(2018)}]{chenorff}%
  \BibitemOpen
  \bibfield  {author} {\bibinfo {author} {\bibfnamefont {A.}~\bibnamefont
  {Dimić}}\ and\ \bibinfo {author} {\bibfnamefont {B.}~\bibnamefont
  {Dakić}},\ }\href {\doibase 10.1038/s41534-017-0055-x} {\bibfield  {journal}
  {\bibinfo  {journal} {npj Quantum Inf.}\ }\textbf {\bibinfo {volume} {4}},\
  \bibinfo {pages} {11} (\bibinfo {year} {2018})}\BibitemShut {NoStop}%
\bibitem [{\citenamefont {Hou}\ \emph {et~al.}(2016)\citenamefont {Hou},
  \citenamefont {Zhu}, \citenamefont {Xiang}, \citenamefont {Li},\ and\
  \citenamefont {Guo}}]{Error}%
  \BibitemOpen
  \bibfield  {author} {\bibinfo {author} {\bibfnamefont {Z.}~\bibnamefont
  {Hou}}, \bibinfo {author} {\bibfnamefont {H.}~\bibnamefont {Zhu}}, \bibinfo
  {author} {\bibfnamefont {G.~Y.}\ \bibnamefont {Xiang}}, \bibinfo {author}
  {\bibfnamefont {C.~F.}\ \bibnamefont {Li}}, \ and\ \bibinfo {author}
  {\bibfnamefont {G.~C.}\ \bibnamefont {Guo}},\ }\href {\doibase
  10.1364/JOSAB.33.001256} {\bibfield  {journal} {\bibinfo  {journal}
  {J.~Opt.~Soc.~Am.~B}\ }\textbf {\bibinfo {volume} {33}},\ \bibinfo {pages}
  {1256} (\bibinfo {year} {2016})}\BibitemShut {NoStop}%
\bibitem [{\citenamefont {Liu}\ \emph {et~al.}(2020)\citenamefont {Liu},
  \citenamefont {Shang}, \citenamefont {Yu},\ and\ \citenamefont
  {Zhang}}]{JWSHANG1}%
  \BibitemOpen
  \bibfield  {author} {\bibinfo {author} {\bibfnamefont {Y.-C.}\ \bibnamefont
  {Liu}}, \bibinfo {author} {\bibfnamefont {J.}~\bibnamefont {Shang}}, \bibinfo
  {author} {\bibfnamefont {X.-D.}\ \bibnamefont {Yu}}, \ and\ \bibinfo {author}
  {\bibfnamefont {X.}~\bibnamefont {Zhang}},\ }\href {\doibase
  10.1103/PhysRevA.101.042315} {\bibfield  {journal} {\bibinfo  {journal}
  {Phys. Rev. A}\ }\textbf {\bibinfo {volume} {101}},\ \bibinfo {pages}
  {042315} (\bibinfo {year} {2020})}\BibitemShut {NoStop}%
\bibitem [{\citenamefont {Zeng}\ \emph {et~al.}(2020)\citenamefont {Zeng},
  \citenamefont {Zhou},\ and\ \citenamefont {Liu}}]{ZengZL20}%
  \BibitemOpen
  \bibfield  {author} {\bibinfo {author} {\bibfnamefont {P.}~\bibnamefont
  {Zeng}}, \bibinfo {author} {\bibfnamefont {Y.}~\bibnamefont {Zhou}}, \ and\
  \bibinfo {author} {\bibfnamefont {Z.}~\bibnamefont {Liu}},\ }\href {\doibase
  10.1103/PhysRevResearch.2.023306} {\bibfield  {journal} {\bibinfo  {journal}
  {Phys. Rev. Research}\ }\textbf {\bibinfo {volume} {2}},\ \bibinfo {pages}
  {023306} (\bibinfo {year} {2020})}\BibitemShut {NoStop}%
\bibitem [{\citenamefont {Pallister}\ \emph {et~al.}(2018)\citenamefont
  {Pallister}, \citenamefont {Linden},\ and\ \citenamefont {Montanaro}}]{QSV2}%
  \BibitemOpen
  \bibfield  {author} {\bibinfo {author} {\bibfnamefont {S.}~\bibnamefont
  {Pallister}}, \bibinfo {author} {\bibfnamefont {N.}~\bibnamefont {Linden}}, \
  and\ \bibinfo {author} {\bibfnamefont {A.}~\bibnamefont {Montanaro}},\ }\href
  {\doibase 10.1103/PhysRevLett.120.170502} {\bibfield  {journal} {\bibinfo
  {journal} {Phys. Rev. Lett.}\ }\textbf {\bibinfo {volume} {120}},\ \bibinfo
  {pages} {170502} (\bibinfo {year} {2018})}\BibitemShut {NoStop}%
\bibitem [{\citenamefont {Je\ifmmode~\check{z}\else \v{z}\fi{}ek}\ \emph
  {et~al.}(2003)\citenamefont {Je\ifmmode~\check{z}\else \v{z}\fi{}ek},
  \citenamefont {Fiur\'a\ifmmode~\check{s}\else \v{s}\fi{}ek},\ and\
  \citenamefont {Hradil}}]{QPT}%
  \BibitemOpen
  \bibfield  {author} {\bibinfo {author} {\bibfnamefont {M.}~\bibnamefont
  {Je\ifmmode~\check{z}\else \v{z}\fi{}ek}}, \bibinfo {author} {\bibfnamefont
  {J.}~\bibnamefont {Fiur\'a\ifmmode~\check{s}\else \v{s}\fi{}ek}}, \ and\
  \bibinfo {author} {\bibfnamefont {Z.}~\bibnamefont {Hradil}},\ }\href
  {\doibase 10.1103/PhysRevA.68.012305} {\bibfield  {journal} {\bibinfo
  {journal} {Phys. Rev. A}\ }\textbf {\bibinfo {volume} {68}},\ \bibinfo
  {pages} {012305} (\bibinfo {year} {2003})}\BibitemShut {NoStop}%
\end{thebibliography}%

\end{document}


\title{Efficient Experimental Verification of
Quantum Gates with Local Operations: Supplemental Material}

\author{Rui-Qi Zhang}
\thanks{These authors contributed equally to this work.}
\affiliation{Key Laboratory of Quantum Information,University of Science and Technology of China, CAS, Hefei 230026, P. R. China}
\affiliation{Synergetic Innovation Center of Quantum Information and Quantum Physics, University of Science and Technology of China, Hefei 230026, P. R. China}
\author{Zhibo Hou}
\thanks{These authors contributed equally to this work.}
\affiliation{Key Laboratory of Quantum Information,University of Science and Technology of China, CAS, Hefei 230026, P. R. China}
\affiliation{Synergetic Innovation Center of Quantum Information and Quantum Physics, University of Science and Technology of China, Hefei 230026, P. R. China}
\author{Jun-Feng Tang}
\affiliation{Key Laboratory of Quantum Information,University of Science and Technology of China, CAS, Hefei 230026, P. R. China}
\affiliation{Synergetic Innovation Center of Quantum Information and Quantum Physics, University of Science and Technology of China, Hefei 230026, P. R. China}
\author{Jiangwei Shang}
\email{jiangwei.shang@bit.edu.cn}
\affiliation{Key Laboratory of Advanced Optoelectronic Quantum Architecture and Measurement of Ministry of Education, School of Physics, Beijing Institute of Technology, Beijing 100081, China}
\author{Huangjun~Zhu}
\email{zhuhuangjun@fudan.edu.cn}
\affiliation{State Key Laboratory of Surface Physics and Department of Physics, Fudan University, Shanghai 200433, China}
\affiliation{Institute for Nanoelectronic Devices and Quantum Computing, Fudan University, Shanghai 200433, China}
\affiliation{Center for Field Theory and Particle Physics, Fudan University, Shanghai 200433, China}
\author{Guo-Yong Xiang}
\email{gyxiang@ustc.edu.cn}
\affiliation{Key Laboratory of Quantum Information,University of Science and Technology of China, CAS, Hefei 230026, P. R. China}
\affiliation{Synergetic Innovation Center of Quantum Information and Quantum Physics, University of Science and Technology of China, Hefei 230026, P. R. China}
\author{Chuan-Feng Li}
\affiliation{Key Laboratory of Quantum Information,University of Science and Technology of China, CAS, Hefei 230026, P. R. China}
\affiliation{Synergetic Innovation Center of Quantum Information and Quantum Physics, University of Science and Technology of China, Hefei 230026, P. R. China}
\author{Guang-Can Guo}
\affiliation{Key Laboratory of Quantum Information,University of Science and Technology of China, CAS, Hefei 230026, P. R. China}
\affiliation{Synergetic Innovation Center of Quantum Information and Quantum Physics, University of Science and Technology of China, Hefei 230026, P. R. China}

\date{\today}

\maketitle


%
\section{Estimation of the confidence Level and infidelity from test results}
In this section we prove Eqs.~(6) and (7) in the main text. Recall that $p_A (\Theta,\epsilon)$ is the maximal passing
probability for quantum gates with infidelity $\epsilon_{A} \geq  \epsilon$ given the process verification operator $\Theta$ \cite{HGZHU1}. By definition, $p_A(\Theta, \epsilon)$ is nonincreasing in $\epsilon$. Note that the passing
probability is upper bounded by $p_A(\Theta, \epsilon)$ whenever the quantum gate under consideration has infidelity $\epsilon_{A} \geq  \epsilon$. Let $\hat{p}_s$ be the passing rate of  a quantum gate in a given experiment with $N$ tests.  The significance level $\delta (\hat{p}_s)$  that the infidelity satisfies  $\epsilon_A<\epsilon$ is defined as the maximal probability that 
the passing rate is larger than $\hat{p}_s$
for any quantum gate that satisfies  $\epsilon_A\geq \epsilon$. 
If  $\hat{p}_s>p_A(\Theta, \epsilon)$, then the Chernoff bound \cite{chenorff} implies that
\begin{equation}\label{eq:significanceUB}
\delta (\hat{p}_s) \le {\rm exp}[-D(\hat{p}_s\| p_A(\Theta, \epsilon))N]\,,
\end{equation}
which is equivalent to  Eq.~(6) in the main text.

To prove Eq.~(7) in the main text, note that, for a balanced verification strategy $\Theta$, the  probability $p_A(\Theta, \epsilon)$ satisfies \cite{HGZHU1}
\begin{equation}
p_A(\Theta, \epsilon) \le 1 - \frac{d+1}{d}\nu(\Theta)\epsilon,
\end{equation}
where $\nu(\Theta)$ is the spectral gap of $\Theta$, that is, the gap between the largest
eigenvalue of $\Theta$ (which is unity) and the second largest eigenvalue of $\Theta$. 
The above equation implies that
\begin{equation}\label{eq:ineq7}
\epsilon \le \frac{d}{d+1} \frac{1 - p_A(\Theta, \epsilon)}{\nu(\Theta)}\,,
\end{equation}
which sets an upper bound for the infidelity
given the maximal  passing probability $p_A(\Theta, \epsilon)$.  Now we
rearrange \eref{eq:significanceUB} to derive
\begin{equation}\label{eq:ineq8}
p_A(\Theta, \epsilon) \ge D^{(-1)}\biggl(\hat{p}_s, \frac{\ln\delta^{-1}}{N}\biggr)\,,
\end{equation}
where $x = D^{(-1)}(\hat{p}_s, y)$ is the inverse function
of $y = D(\hat{p}_s\|x)$ with domain $x\in (0, \hat{p}_s]$.  Note that
$D(\hat{p}_s\|x)$ is strictly decreasing in $x\in (0, \hat{p}_s]$, so 
$D^{(-1)}(\hat{p}_s, y)$ is well defined for $y\geq0$.
Combining \esref{eq:ineq7} and   \eqref{eq:ineq8}, we can deduce that 
\begin{equation}
\epsilon_{A} \le \frac{d}{d+1} \frac{1-D^{(-1)}\bigl(\hat{p}_s, \frac{\ln\delta^{-1}}{N}\bigr)}{\nu(\Theta)}\,,
\end{equation}
which confirms Eq.~(7) in the main text.

\section{Experimental implementations}
In this section we show that the two state-preparation modules in Fig.~2 in the main text
can prepare arbitrary product states of two- and three-qubit systems, respectively.
In addition, the two measurement modules in Fig.~2  can perform
arbitrary local projective measurements on two- and three-qubit systems, respectively.

\subsection{Preparation of two-qubit product states}
\begin{figure}[t]
	\center{\includegraphics[scale=0.35]{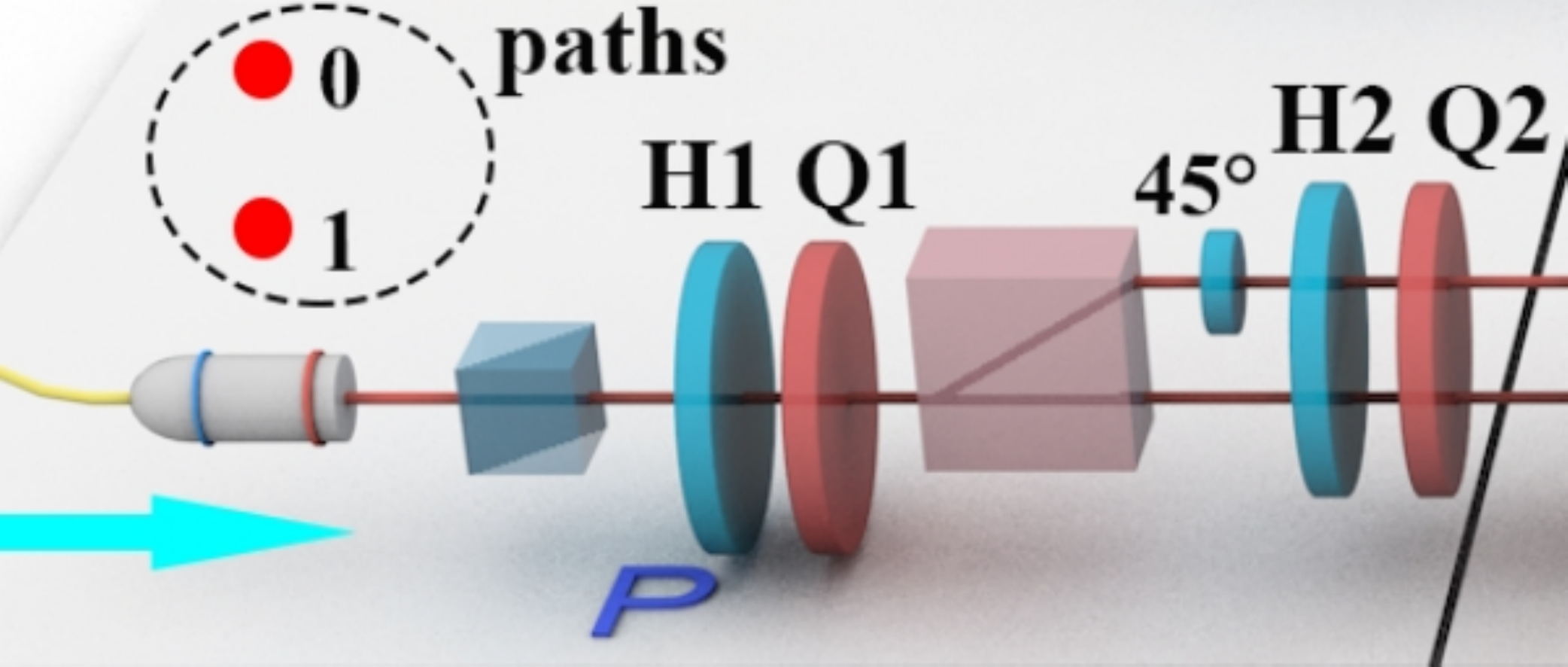}}
	\caption{\label{fig:preparation_2qubit}
		The state-preparation module for the verification of the {\sc cnot} gate.}
\end{figure}

The state-preparation module for the verification of the {\sc cnot} gate is reproduced in Fig.~\ref{fig:preparation_2qubit}. 
Suppose the angles of the optical axes of H1, Q1, H2, and Q2 in the figure are
$h_1$, $q_1$, $h_2$, and $q_2$ (with respect to the horizontal direction), respectively. The input
heralded photon is initially prepared in the state $|1\rangle\otimes|H\rangle$, which
is transformed into the following state
\begin{align}
    &|1\rangle \otimes
    \bigl[\cos(q_1)\cos(q_1 - 2h_1) + \rmi \sin(q_1)\sin(q_1 - 2h_1)\bigr]|H\rangle\nonumber\\
    +&|1\rangle \otimes
    \bigl[\sin(q_1)\cos(q_1 - 2h_1) - \rmi \cos(q_1)\sin(q_1 - 2h_1)\bigr]|V\rangle
\end{align}
by H1 and Q1. The BD after Q1 coherently routes the heralded photon to paths $0$ and $1$
according to the polarization state of the heralded photon, which yields the state
\begin{align}
    &|0\rangle \otimes
    \bigl[\sin(q_1)\cos(q_1 - 2h_1) - \rmi \cos(q_1)\sin(q_1 - 2h_1)\bigr]|V\rangle\nonumber \\
    +&|1\rangle \otimes
    \bigl[\cos(q_1)\cos(q_1 - 2h_1) + \rmi \sin(q_1)\sin(q_1 - 2h_1)\bigr]|H\rangle\,.
\end{align}
Then H2 and Q3 together with a $45^\circ$ HWP transform the photon state into
\begin{equation}
    (a_0|0\rangle + a_1|1\rangle) \otimes (b_0|H\rangle + b_1|V\rangle)\,,
\end{equation}
where
\begin{equation}\label{eq:state_preparation_param}
\begin{aligned}
    a_0 := \sin(q_1)\cos(q_1 - 2h_1) - \rmi \cos(q_1)\sin(q_1 - 2h_1)\,,\\
    a_1 := \cos(q_1)\cos(q_1 - 2h_1) + \rmi \sin(q_1)\sin(q_1 - 2h_1)\,,\\
    b_0 := \cos(q_2)\cos(q_2 - 2h_2) + \rmi \sin(q_2)\sin(q_2 - 2h_2)\,,\\
    b_1 := \sin(q_2)\cos(q_2 - 2h_2) - \rmi \cos(q_2)\sin(q_2 - 2h_2)\,.
\end{aligned}
\end{equation}
Hence, to prepare the state in Eq.~(8) in the main text, 
we need to choose appropriate values of the parameters $h_1$, $q_1$, $h_2$, and $q_2$ 
to satisfy the conditions in \eref{eq:state_preparation_param}.

\subsection{Preparation of three-qubit product states}
\begin{figure}[t]
	\center{\includegraphics[scale=0.3]{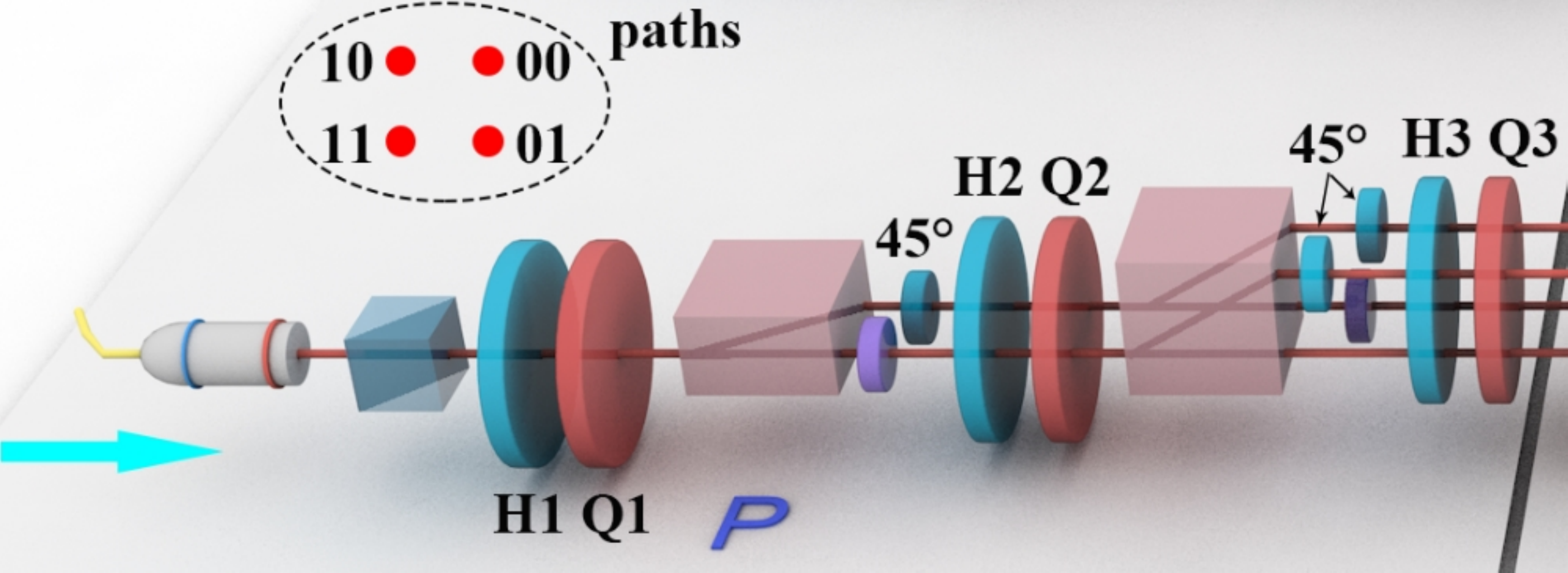}}
	\caption{\label{fig:preparation_3qubit}
		The state-preparation module for the verification of the Toffoli gate.}
\end{figure}

The state-preparation module for the verification of the Toffoli gate is reproduced in Fig.~\ref{fig:preparation_3qubit}.
Suppose the angles of the optical axes of H1, Q1, H2, Q2, H3, and Q3 in the figure are
$h_1$, $q_1$, $h_2$, $q_2$, $h_3$, and $q_3$ (with respect to the horizontal direction),
respectively. The input heralded photon is initially prepared in the state
$|11\rangle\otimes|H\rangle$, which is turned into the following state
\begin{align}
    &|11\rangle \otimes
    \bigl[\cos(q_1)\cos(q_1 - 2h_1) + \rmi \sin(q_1)\sin(q_1 - 2h_1)\bigr]|H\rangle\nonumber \\
    +&|11\rangle \otimes
    \bigl[\sin(q_1)\cos(q_1 - 2h_1) - \rmi \cos(q_1)\sin(q_1 - 2h_1)\bigr]|V\rangle
\end{align}
by H1 and Q1. The BD after Q1 coherently routes the heralded photon to paths $11$ and $01$
according to the polarization state of the heralded photon, which yields the state
\begin{align}
    &|01\rangle \otimes
    \bigl[\sin(q_1)\cos(q_1 - 2h_1) - \rmi \cos(q_1)\sin(q_1 - 2h_1)\bigr]|V\rangle\nonumber \\
    +&|11\rangle \otimes
    \bigl[\cos(q_1)\cos(q_1 - 2h_1) + \rmi \sin(q_1)\sin(q_1 - 2h_1)\bigr]|H\rangle\,.
\end{align}
Then a $45^\circ$ HWP (which flips the photon's polarization) together with a K9 plate (which compensates the path-length difference) transforms the photon state into
\begin{equation}
    (a_0'|0\rangle + a_1'|1\rangle) \otimes |1\rangle \otimes |H\rangle\,,
\end{equation}
where
\begin{equation}\label{eq:state_preparation2_1}
\begin{aligned}
    &a_0' := \sin(q_1)\cos(q_1 - 2h_1) - \rmi \cos(q_1)\sin(q_1 - 2h_1)\,, \\
    &a_1' := \cos(q_1)\cos(q_1 - 2h_1) + \rmi \sin(q_1)\sin(q_1 - 2h_1)\,.
\end{aligned}
\end{equation}
This initializes the first qubit.

Next,  H2 and Q2 transform the photon state into
\begin{align}
    (a_0'|0\rangle + a_1'|1\rangle) \otimes |1\rangle
    \otimes \bigl\{&[\cos(q_2)\cos(q_2 - 2h_2) + \rmi \sin(q_2)\sin(q_2 - 2h_2)]|H\rangle\nonumber \\
    +&\bigl[\sin(q_2)\cos(q_2 - 2h_2) - \rmi \cos(q_2)\sin(q_2 - 2h_2)\bigr]|V\rangle\bigr\}\,.
\end{align}
Then  the second BD together with two $45^\circ$ HWPs and a K9 plate
transform the state of the heralded photon into
\begin{equation}
    (a_0'|0\rangle + a_1'|1\rangle) \otimes (b_0'|0\rangle + b_1'|1\rangle) \otimes |H\rangle\,,
\end{equation}
where
\begin{equation}\label{eq:state_preparation2_2}
\begin{aligned}
    &b_0' := \sin(q_2)\cos(q_2 - 2h_2) - \rmi \cos(q_2)\sin(q_2 - 2h_2)\,, \\
    &b_1' := \cos(q_2)\cos(q_2 - 2h_2) + \rmi \sin(q_2)\sin(q_2 - 2h_2)\,.
\end{aligned}
\end{equation}
This initializes the second qubit.

Finally, H3 and Q3 transform the photon state into
\begin{equation}
    (a_0'|0\rangle + a_1'|1\rangle) \otimes (b_0'|0\rangle + b_1'|1\rangle)
    \otimes (c_0'|H\rangle + c_1'|V\rangle)\,,
\end{equation}
where
\begin{equation}\label{eq:state_preparation2_3}
\begin{aligned}
    &c_0' := \sin(q_3)\cos(q_3 - 2h_3) - \rmi \cos(q_3)\sin(q_3 - 2h_3)\,, \\
    &c_1' := \cos(q_3)\cos(q_3 - 2h_3) + \rmi \sin(q_3)\sin(q_3 - 2h_3)\,.
\end{aligned}
\end{equation}
This initializes the third qubit. Hence, to prepare the state
in Eq.~(9) in the main text, we need to choose
proper values of the parameters $h_1$, $q_1$, $h_2$, $q_2$, $h_3$, and $q_3$ to satisfy the conditions in
\esref{eq:state_preparation2_1}, \eqref{eq:state_preparation2_2}, and
\eqref{eq:state_preparation2_3} simultaneously.

\begin{figure}[t]
	\center{\includegraphics[scale=0.5]{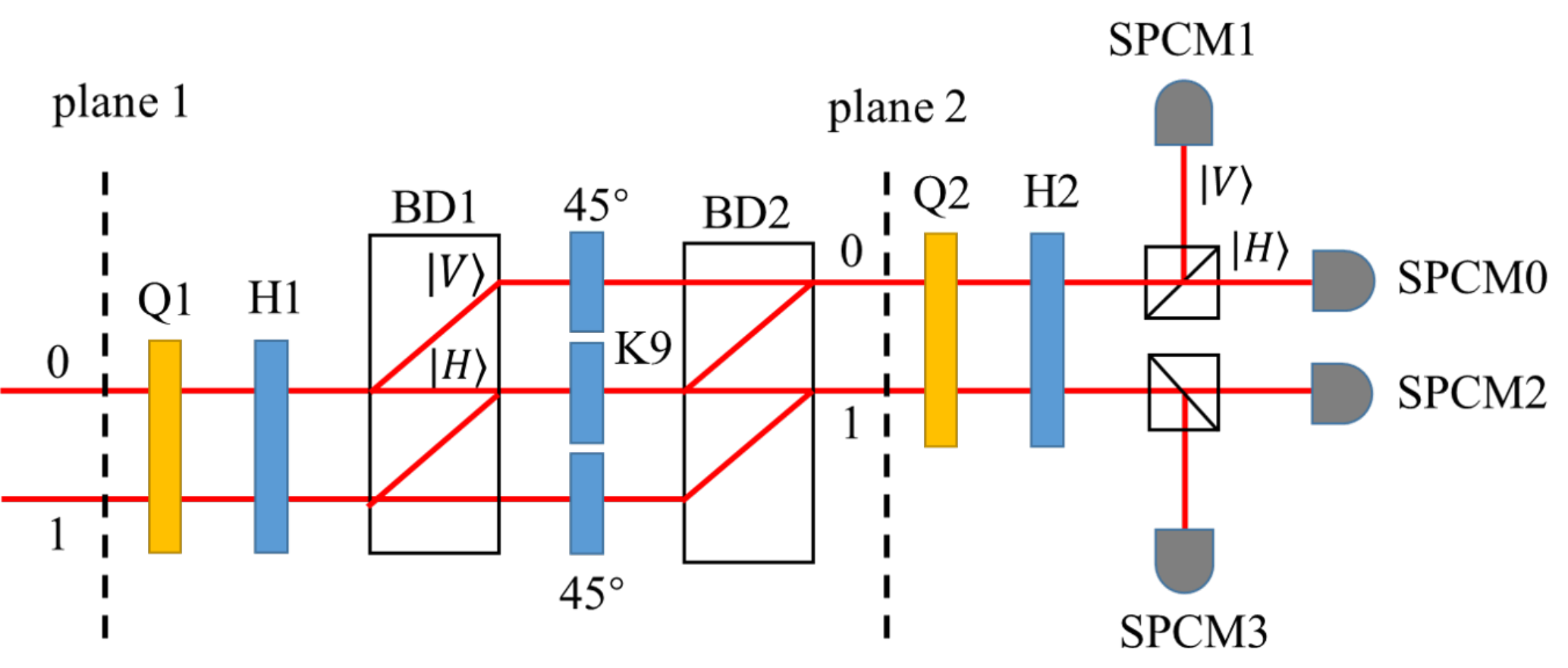}}
	\caption{\label{fig:measurement_2qubit}
    The measurement module employed for the verification of the {\sc cnot} gate. The PAS in Fig.~2 in the main text is unfolded here to better illustrate the structure of the measurement module.}
\end{figure}
\subsection{Implementation of local projective measurements on a two-qubit system}
Here we show that the measurement module employed for the verification of the {\sc cnot} gate shown in Fig. 2 in the main text as reproduced in Fig.~\ref{fig:measurement_2qubit}  can realize arbitrary local projective
measurements on the output two-qubit state of the quantum-gate module.

Suppose we want to perform projective measurements onto orthonormal bases
$\{|\varphi^{(0)}_0\rangle, |\varphi^{(0)}_1\rangle\}$ and
$\{|\varphi^{(1)}_0\rangle, |\varphi^{(1)}_1\rangle\}$ for
the path qubit and polarization qubit, respectively.
To simplify the notation, here we temporarily use $|0\rangle$ ($|1\rangle$) to
refer to "horizontal polarization" ("vertical polarization") for the
polarization qubit. Suppose the photon state at plane 1 has the form
\begin{equation}
    |\Psi\rangle = \sum_{i,j \in \{0, 1\}}
    {a_{ij}|\varphi^{(0)}_i\rangle|\varphi^{(1)}_j\rangle}\,.
\end{equation}
We first choose the angles of the optical axes of Q1 and H1 in
Fig.~\ref{fig:measurement_2qubit} so
as to implement the transformation
$|\varphi^{(1)}_0\rangle \to |0\rangle,\ |\varphi^{(1)}_1\rangle \to |1\rangle$
on the polarization qubit (see the Supplemental Material of Ref. \cite{Error} on how
to calculate the angles), which turns $|\Psi\rangle$ into
\begin{equation}
    |\Psi'\rangle = \sum_{i,j \in \{0, 1\}}
    {a_{ij}|\varphi^{(0)}_i\rangle|j\rangle}\,.
\end{equation}
Then the two BDs split and re-combine the two light beams, which turns $|\Psi'\rangle$ into
\begin{equation}
    |\Psi''\rangle = \sum_{i,j \in \{0, 1\}}
    {a_{j(1-i)}|i\rangle|\varphi^{(0)}_j\rangle}
\end{equation}
at plane 2.

Next, we set the angles of the optical axes of Q2 and H2 so as to realize the
transformation $|\varphi^{(0)}_0\rangle \to |0\rangle, \ |\varphi^{(0)}_1\rangle \to |1\rangle$,
which turns $|\Psi''\rangle$ into
\begin{equation}
    |\Psi'''\rangle = \sum_{i,j \in \{0, 1\}}
    {a_{j(1-i)}|i\rangle|j\rangle}\,.
\end{equation}
Now, the probabilities that the heralded photon is detected by SPCM 0 to 3 are given by
$|a_{01}|^2$, $|a_{11}|^2$, $|a_{00}|^2$, $|a_{10}|^2$, respectively.
In this way, we can realize the projective measurement onto the product basis
\begin{equation}
\left\{|\varphi^{(0)}_0\rangle|\varphi^{(1)}_0\rangle, |\varphi^{(0)}_0\rangle|\varphi^{(1)}_1\rangle, |\varphi^{(0)}_1\rangle|\varphi^{(1)}_0\rangle, |\varphi^{(0)}_1\rangle|\varphi^{(1)}_1\rangle\right\}.
\end{equation}

\begin{figure}[t]
	\center{\includegraphics[scale=0.68]{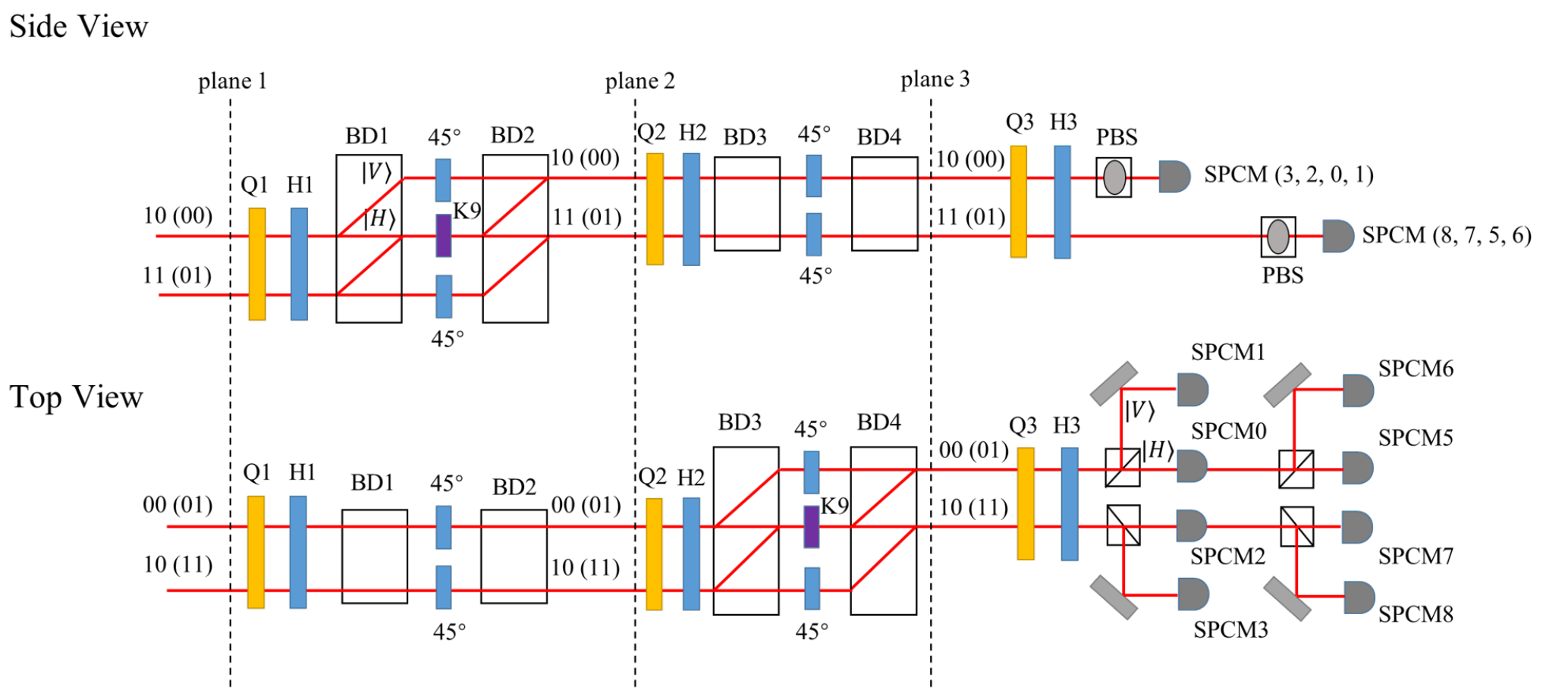}}
	\caption{\label{fig:measurement_3qubit}
    The measurement module employed for the verification of the Toffoli gate. Since the trajectory
of the photon cannot be described in a plane, we provide both 
the side view and the top view.
The PAS  in Fig.~2 in the main text
is unfolded here to better illustrate the structure
of the measurement module. The mirror right ahead of
the PAS in Fig.~2 is removed in this figure.}
\end{figure}
\subsection{Implementation of local projective measurements on a three-qubit system}
Here we show that the measurement module employed for the verification of the Toffoli gate
shown in Fig.~2 in the main text as reproduced in Fig.~\ref{fig:measurement_3qubit}  can realize arbitrary local projective
measurements on the output three-qubit states of the quantum-gate module.

Suppose we want to perform projective measurements onto the three orthonormal bases
$\{|\psi^{(0)}_0\rangle, |\psi^{(0)}_1\rangle\}$,
$\{|\psi^{(1)}_0\rangle, |\psi^{(1)}_1\rangle\}$,
and $\{|\psi^{(2)}_0\rangle, |\psi^{(2)}_1\rangle\}$ for
the two path qubits and one polarization qubit, respectively.
To simplify the notation, here we temporarily use $|0\rangle$ ($|1\rangle$) to refer to "horizontal polarization" ("vertical polarization") for the
polarization qubit. Suppose the photon state at plane 1 has the form
\begin{equation}
    |\Psi_0\rangle =
    \sum_{i,j,k \in \{0, 1\}}
    {a_{ijk}|\psi^{(0)}_i\rangle|\psi^{(1)}_j\rangle|\psi^{(2)}_k\rangle}\,.
\end{equation}
We first choose the angles of the optical axes of Q1 and H1 in
Fig.~\ref{fig:measurement_3qubit} so
as to implement the transformation $|\psi^{(2)}_0\rangle \to |0\rangle,
|\psi^{(2)}_1\rangle \to |1\rangle$
on the polarization qubit, which turns $|\Psi_0\rangle$ into
\begin{equation}
        |\Psi_1\rangle =
        \sum_{i,j,k \in \{0, 1\}}
        {a_{ijk}|\psi^{(0)}_i\rangle|\psi^{(1)}_j\rangle|k\rangle}\,.
\end{equation}
Then BD1 and BD2 split and re-combine the two light beams, which turns $|\Psi_1\rangle$ into
\begin{equation}
    |\Psi_2\rangle =
        \sum_{i,j,k \in \{0, 1\}}
        {a_{ik(1-j)}|\psi^{(0)}_i\rangle|j\rangle|\psi^{(1)}_k\rangle}
\end{equation}
at plane 2.

Next, we set the angles of the optical axes of Q2 and H2 so as to realize the
transformation $|\psi^{(1)}_0\rangle \to |0\rangle,
|\psi^{(1)}_1\rangle \to |1\rangle$ on the polarization qubit, which
turns $|\Psi_2\rangle$ into
\begin{equation}
    |\Psi_3\rangle =
        \sum_{i,j,k \in \{0, 1\}}
        {a_{ik(1-j)}|\psi^{(0)}_i\rangle|j\rangle|k\rangle}\,.
\end{equation}
Then BD3 and BD4 split and re-combine the two light beams, which turns
$|\Psi_3\rangle$ into
\begin{equation}
    |\Psi_4\rangle =
        \sum_{i,j,k \in \{0, 1\}}
        {a_{k(1-j)(1-i)}|i\rangle|j\rangle|\psi^{(0)}_k\rangle}
\end{equation}
at plane 3.

Finally, we set the angles of the optical axes of Q3 and H3 so as to realize the
transformation $|\psi^{(0)}_0\rangle \to |0\rangle,
|\psi^{(0)}_1\rangle \to |1\rangle$ on the polarization qubit, which turns $|\Psi_4\rangle$ into
\begin{equation}
    |\Psi_5\rangle =
        \sum_{i,j,k \in \{0, 1\}}
        {a_{k(1-j)(1-i)}|i\rangle|j\rangle|k\rangle}\,.
\end{equation}
Now, the probabilities that the heralded photon is detected by  SPCMs 0 to 7 are given by
$|a_{011}|^2$, $|a_{111}|^2$, $|a_{001}|^2$, $|a_{101}|^2$, $|a_{010}|^2$,
$|a_{111}|^2$, $|a_{000}|^2$, $|a_{100}|^2$, respectively. 
In this way, we can realize the projective measurement onto the three-qubit product basis
\begin{align}
\Bigl\{&|\psi^{(0)}_0\rangle|\psi^{(1)}_0\rangle|\psi^{(2)}_0\rangle,\,
    |\psi^{(0)}_0\rangle|\psi^{(1)}_0\rangle|\psi^{(2)}_1\rangle,\,
    |\psi^{(0)}_0\rangle|\psi^{(1)}_1\rangle|\psi^{(2)}_0\rangle,\,
    |\psi^{(0)}_0\rangle|\psi^{(1)}_1\rangle|\psi^{(2)}_1\rangle,\,\nonumber\\
    &|\psi^{(0)}_1\rangle|\psi^{(1)}_0\rangle|\psi^{(2)}_0\rangle,\,
    |\psi^{(0)}_1\rangle|\psi^{(1)}_0\rangle|\psi^{(2)}_1\rangle,\,
    |\psi^{(0)}_1\rangle|\psi^{(1)}_1\rangle|\psi^{(2)}_0\rangle,\,
    |\psi^{(0)}_1\rangle|\psi^{(1)}_1\rangle|\psi^{(2)}_1\rangle
    \Bigr\}.
\end{align}

\section{Protocols for verifying the {\sc cnot} and Toffoli gates}
The procedure for constructing a verification protocol for a unitary
transformation $\mathcal U$ consists of  two steps \cite{JWSHANG1,HGZHU1,ZengZL20}. First, we choose an
ensemble of test states $\{\rho_j, p_j\}_j$, where each  $\rho_j=|\psi_j\rangle\langle\psi_j|$ is a pure state
and $p_j$ is the probability of picking $\rho_j$. Next, we construct a verification strategy for each output state $\mathcal U(\rho_j)$ according to the framework of quantum state verification (QSV) \cite{QSV2}. A verification strategy for $\mathcal U(\rho_j)$ consists of a number of projective tests with test projectors $M_{l}^{(j)}$, which are chosen with probabilities $p_{l|j}$ given the test state $\rho_j$. The resulting verification operator for the state $\mathcal U(\rho_j)$ reads $\Omega_j =
\sum_lp_{l|j}M_l^{(j)}$. 
By contrast,  the process
verification operator for $\mathcal U$ reads
\begin{equation}
\Theta := d\sum_{j}{p_j \mathcal{U}^{-1}\biggl(\sum_{l}{p_{l|j}
		M_{l}^{(j)}}\biggr)\otimes\rho_j^{*}} = d\sum_{j}{p_j
	\mathcal{U}^{-1}(\Omega_j)\otimes\rho_j^{*}}.
\end{equation}
When the set $\{\rho_j, p_j\}_j$ of test states is balanced, which means the condition $\sum_j p_j\rho_j=\openone/d$ holds, the efficiency of the verification protocol is mainly determined by the spectral gap $\nu(\Theta)$ as shown in  Eq.~(5) in the main text. A larger spectral gap usually means a higher efficiency.

\begin{table}[htbp]
	\renewcommand\arraystretch{1.3}	
	\caption{\label{table:test_state}
		Test states for the verification of the {\sc cnot} gate. Here $|H\rangle$ and
		$|V\rangle$ denote the eigenstates of $Z$ with eigenvalues
		$1$ and $-1$, respectively; $|D\rangle, |A\rangle, |R\rangle, |L\rangle$ are defined as
		$|D\rangle = \frac{|H\rangle + |V\rangle}{\sqrt{2}}$,
		$|A\rangle = \frac{|H\rangle - |V\rangle}{\sqrt{2}}$,
		$|R\rangle = \frac{|H\rangle + \rmi|V\rangle}{\sqrt{2}}$,
		$|L\rangle = \frac{|H\rangle - \rmi|V\rangle}{\sqrt{2}}$.}
	\begin{center}
		\setlength{\tabcolsep}{7mm}{\begin{tabular}{|l|l|l|l|l|l|}
				\hline
				\multicolumn{1}{|c|}{bases} & \multicolumn{4}{c|}{states}                               & index $j$  \\ \hline
				$Z\otimes Z$                & $|HH\rangle$ & $|HV\rangle$ & $|VH\rangle$ & $|VV\rangle$ & 1 $\sim$4  \\ \hline
				$X\otimes X$                & $|DD\rangle$ & $|DA\rangle$ & $|AD\rangle$ & $|AA\rangle$ & 5 $\sim$8  \\ \hline
				$Y\otimes Y$                & $|RR\rangle$ & $|RL\rangle$ & $|LR\rangle$ & $|LL\rangle$ & 9 $\sim$12 \\ \hline
		\end{tabular}}
	\end{center}
\end{table}
\subsection{Verification protocol for the {\sc cnot} gate}
The matrix representation of the {\sc cnot} gate in the computational basis is
\begin{equation}
    CX =
    \left[ \begin{array}{cccc}
        1 & 0 & 0 & 0\\
        0 & 1 & 0 & 0\\
        0 & 0 & 0 & 1\\
        0 & 0 & 1 & 0\\
    \end{array}
    \right ]\!.
\end{equation}
As test states, we choose the product eigenstates of the Pauli operators
$Z\otimes Z$, $X\otimes X$, and $Y\otimes Y$,  and each of them is selected with an equal probability of  $1/12$. These test states are labeled as  $|\psi_j\rangle, j = 1,...,12$, where the
correspondence between the index $j$ and the test state is shown in \tref{table:test_state}.

After the action of the {\sc cnot} gate, each ideal output state $CX|\psi_j\rangle$ for $j = 1,...,8$
is a product state and an eigenstate of a Pauli operator, and so can be verified by one measurement setting based on a Pauli measurement; the corresponding verification operator coincides with the projector onto $CX|\psi_j\rangle$, that is, 
\begin{align}
\Omega_j = CX|\psi_j\rangle\langle\psi_j|CX^{\dagger},\quad j = 1,...,8.
\end{align}
By contrast, each  ideal output state $CX|\psi_j\rangle$ for $j = 9,...,12$ is a Bell state and can be verified by three measurement settings based on Pauli measurements \cite{QSV2}. To be specific, the  verification operator $\Omega_j$  has the form 
\begin{equation}\label{eq:qsv_2qubit}
\Omega_j = \frac{1}{3}\Bigl(P^{+}_{(-1)^{b_0+b_1+1}XZ}
+ P^{+}_{(-1)^{b_1}YX} + P^{+}_{(-1)^{b_0}ZY}\Bigr)=\frac{1}{3}\bigl(2CX|\psi_j\rangle\langle\psi_j|CX^{\dagger}+\openone\bigr),\;\; j = 9,...,12,
\end{equation}
where $b_1b_0$ is the binary representation of the  number $j-9$,
and $P^{+}_{O_1O_2}$ denotes the projector onto the eigenspace of   $O_1\otimes O_2$ associated with eigenvalue 1. The spectral gap of the resulting process verification operator $\Theta$ reads
\begin{equation}
\nu(\Theta) = \frac{5}{9}.
\end{equation}

\begin{table}[H]
	\renewcommand\arraystretch{1.3}
	\caption{\label{table:test_state_toffoli}
		Test states for the verification of the Toffoli gate. Here $|H\rangle$ and
		$|V\rangle$ denote the eigenstates of $Z$ with eigenvalues
		$1$ and $-1$, respectively;  $|D\rangle, |A\rangle$ are defined as
		$|D\rangle = \frac{|H\rangle + |V\rangle}{\sqrt{2}}$,
		$|A\rangle = \frac{|H\rangle - |V\rangle}{\sqrt{2}}$.}
	\begin{center}
		\setlength{\tabcolsep}{1.7mm}{
			\begin{tabular}{|l|l|l|l|l|l|l|l|l|l|}
				\hline
				\multicolumn{1}{|c|}{bases} & \multicolumn{8}{c|}{states}                                                                                                                & index $j$  \\ \hline
				$Z\otimes Z \otimes X$      & $|HHD\rangle$ & $|HHA\rangle$ & $|HVD\rangle$ & $|HVA\rangle$ & $|VHD\rangle$ & $|VHA\rangle$ & $|VVD\rangle$ & $|VVA\rangle$ & 1 $\sim$8  \\ \hline
				$X\otimes X \otimes Z$      & $|DDH\rangle$ & $|DDV\rangle$ & $|DAH\rangle$ & $|DAV\rangle$ & $|ADH\rangle$ & $|ADV\rangle$ & $|AAH\rangle$ & $|AAV\rangle$              & 9 $\sim$16 \\ \hline
		\end{tabular}}
	\end{center}
	
\end{table}
\subsection{Verification protocol for the Toffoli gate}
The matrix representation of the Toffoli gate in the computational bases is
\begin{equation}
    C^2X =
    \left[ \begin{array}{cccccccc}
        1 & 0 & 0 & 0 & 0 & 0 & 0 & 0\\
        0 & 1 & 0 & 0 & 0 & 0 & 0 & 0\\
        0 & 0 & 1 & 0 & 0 & 0 & 0 & 0\\
        0 & 0 & 0 & 1 & 0 & 0 & 0 & 0\\
        0 & 0 & 0 & 0 & 1 & 0 & 0 & 0\\
        0 & 0 & 0 & 0 & 0 & 1 & 0 & 0\\
        0 & 0 & 0 & 0 & 0 & 0 & 0 & 1\\
        0 & 0 & 0 & 0 & 0 & 0 & 1 & 0\\
    \end{array}
    \right ]\!.
\end{equation}
As test states, we choose the product eigenstates of the Pauli
operators $Z\otimes Z\otimes X$ and $X\otimes X\otimes Z$, and each of them is
selected with an equal probability of $1/16$.
These test states are labeled as  $|\psi_j\rangle, j = 1,...,16$,
where the correspondence between the index $j$ and the test state is shown
in \tref{table:test_state_toffoli}.

After the action of the Toffoli gate, each ideal output state $C^2X|\psi_j\rangle$ for $j = 1,...,8$ is still a product state and an eigenstate of a Pauli operator and so can be verified by one measurement setting based on a Pauli measurement; the resulting verification operator coincides with the projector onto $C^2X|\psi_j\rangle$, that is,
\begin{align}
\Omega_j =
C^2X|\psi_j\rangle\langle\psi_j|(C^2X)^{\dagger},\quad j = 1,...,8.
\end{align}
By contrast, each  ideal output state $C^2X|\psi_j\rangle$ for $j = 9,...,16$ is  a three-qubit hypergraph state up to some local unitary transformation and can be verified by a variant of the cover protocol proposed in Ref.~\cite{QSV2}. To be specific, the verification strategy consists of three tests based on Pauli measurements chosen with probability $1/3$ each. The resulting verification operator
reads
\begin{equation}\label{eq:qsv_3qubit}
\Omega_j = \frac{1}{3}\left[ f_1((-1)^{b_2}X, Z, X)
+ f_2(Z, (-1)^{b_1}X, X)
+ f_3(Z, Z, (-1)^{b_0}Z)
\right ],\quad  j = 9,...,16,
\end{equation}
where $b_2b_1b_0$ is the binary representation of the number $j-9$. Each test operation in \eref{eq:qsv_3qubit} has the form
\begin{align}\label{eq:f_k}
f_k(O_1, O_2, O_3) := |O_k^{+}\rangle\langle O_k^{+}|\otimes\bigl(\openone -
|O_s^{-}O_t^{-}\rangle\langle O_s^{-}O_t^{-}|\bigr)
+ |O_k^{-}\rangle\langle O_k^{-}|\otimes
|O_s^{-}O_t^{-}\rangle\langle O_s^{-}O_t^{-}|,
\end{align}
 where $k, s, t \in \{1,2,3\}$ are different from each other, $O_i$ for  $i=1,2,3$
are Hermitian operators acting on the $i$th qubit, and $|O_i^{+}\rangle$
($|O_i^{-}\rangle$) is
the eigenstate of $O_i$ with eigenvalue $1$ ($-1$). 
The test $f_k(O_1, O_2, O_3)$ can be realized as follows: 
measure $O_1$, $O_2$, and 
$O_3$ on the three qubits, respectively, and accept the 
state if either (a) the measurement result
on the $k$th qubit is $1$ and the measurement results on the other
two qubits are not $\{-1, -1\}$; or (b) the measurement result
on the $k$th qubit is $-1$ and the measurement results on the other
two qubits are $\{-1, -1\}$.  The spectral gap of the resulting process verification operator $\Theta$ reads
\begin{equation}
\nu(\Theta) = \frac{1}{6}\,.
\end{equation}

%

\begin{figure}[t]
	\center{\includegraphics[scale=0.4]{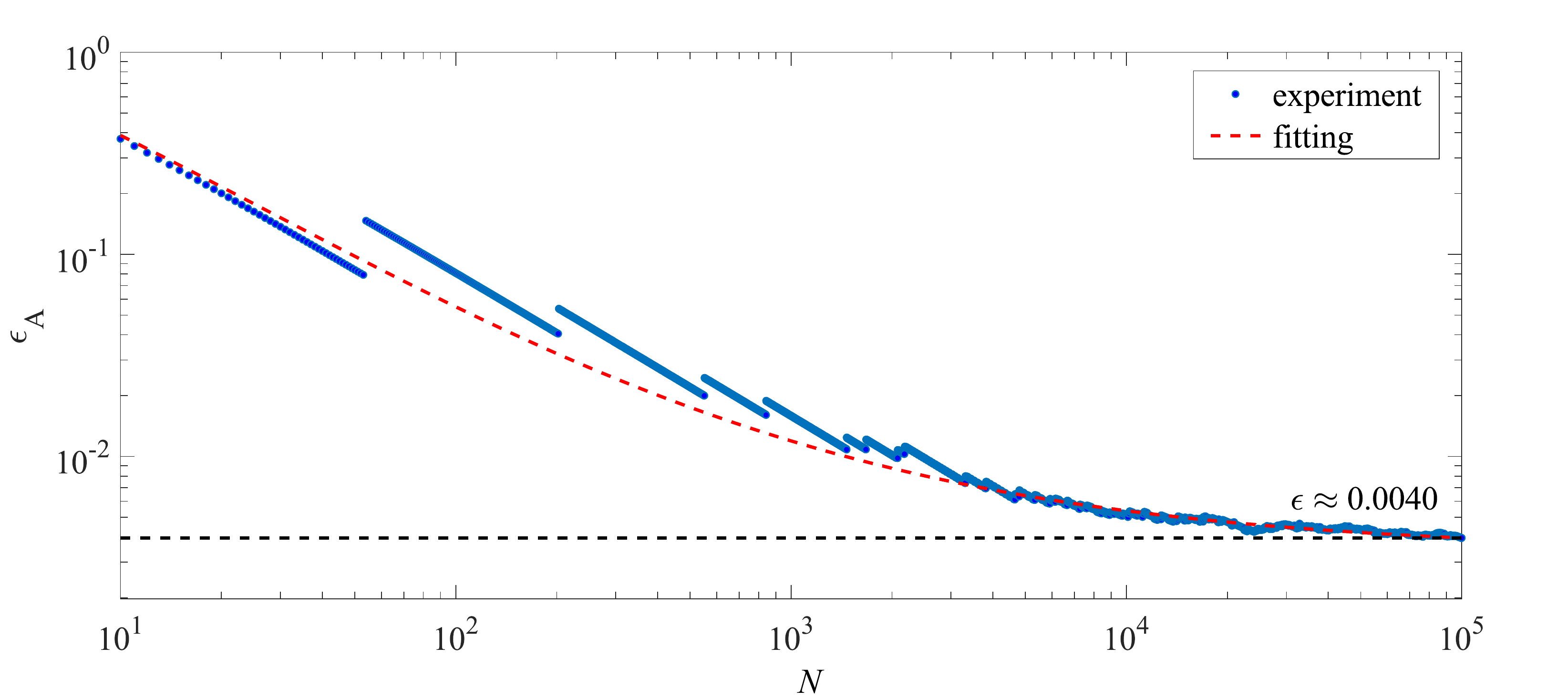}}
	\caption{\label{fig:infidelity_for_CNOT}
Infidelity estimation of the {\sc cnot} gate based on  QGV. Here the confidence level
$1-\delta$  is set to be $95\%$.		
    }
\end{figure}
\begin{figure}[t]
	\center{\includegraphics[scale=0.4]{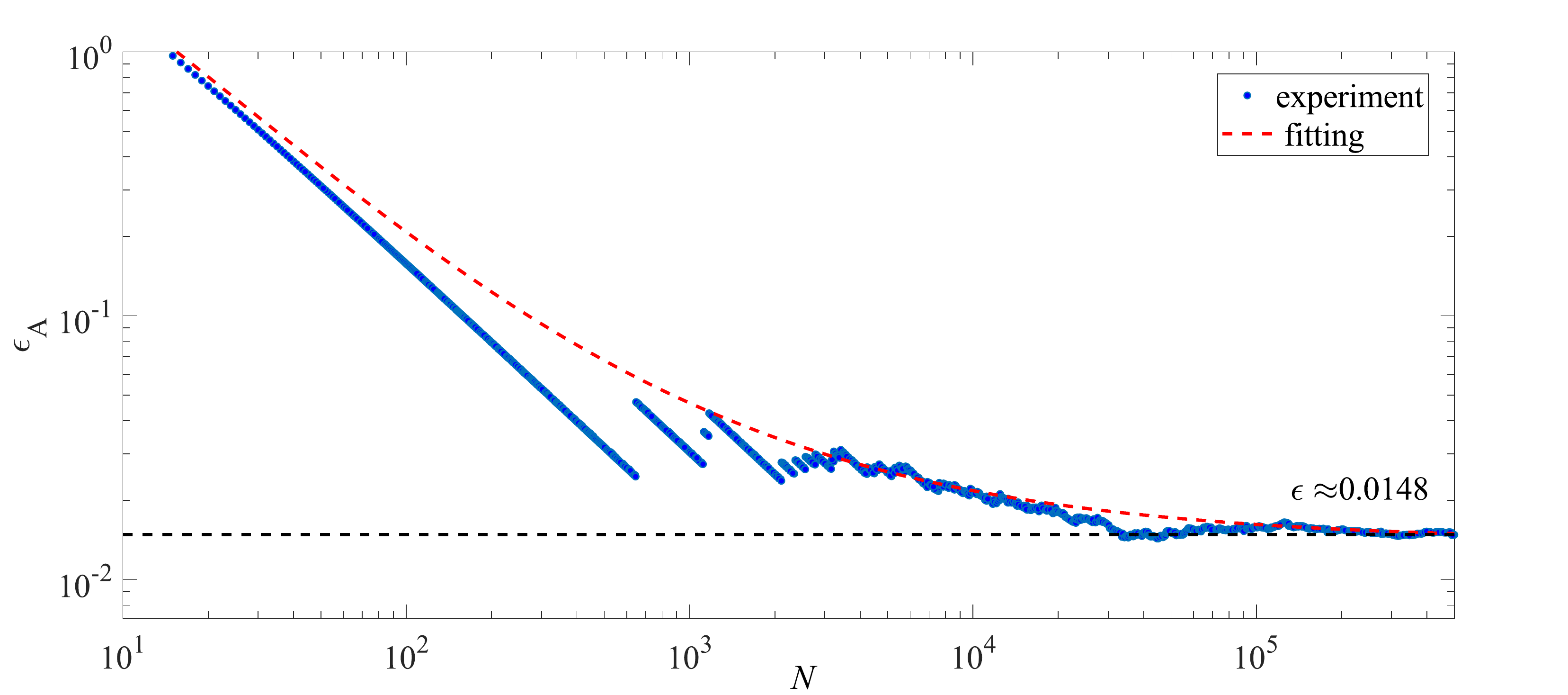}}
	\caption{\label{fig:infidelity_for_Toffoli}
Infidelity estimation of the Toffoli gate based on  QGV. Here the confidence level
$1-\delta$  is set to be $95\%$.		
    }
\end{figure}
\section{Infidelity estimation of the {\sc cnot} gate and the Toffoli gate}
To understand the large-sample limit of practical QGV, we also perform QGV of the {\sc cnot} gate  with $10^5$ tests and QGV of the Toffoli gate
with $5\times10^5$ tests. Equation~(7) in the main text is employed to estimate (upper bounds for) the infidelities of the two quantum gates, and the results are shown in Figs.~\ref{fig:infidelity_for_CNOT} and \ref{fig:infidelity_for_Toffoli}, respectively.
The infidelity of the {\sc cnot} gate
converges to $0.004$ after $10^5$ tests, which is quite close to
the  result $0.003$ determined by QPT. The infidelity of the Toffoli gate converges
to $0.015$ after $2\times10^5$ tests.

\begin{figure}[t]
	\center{\includegraphics[scale=0.6]{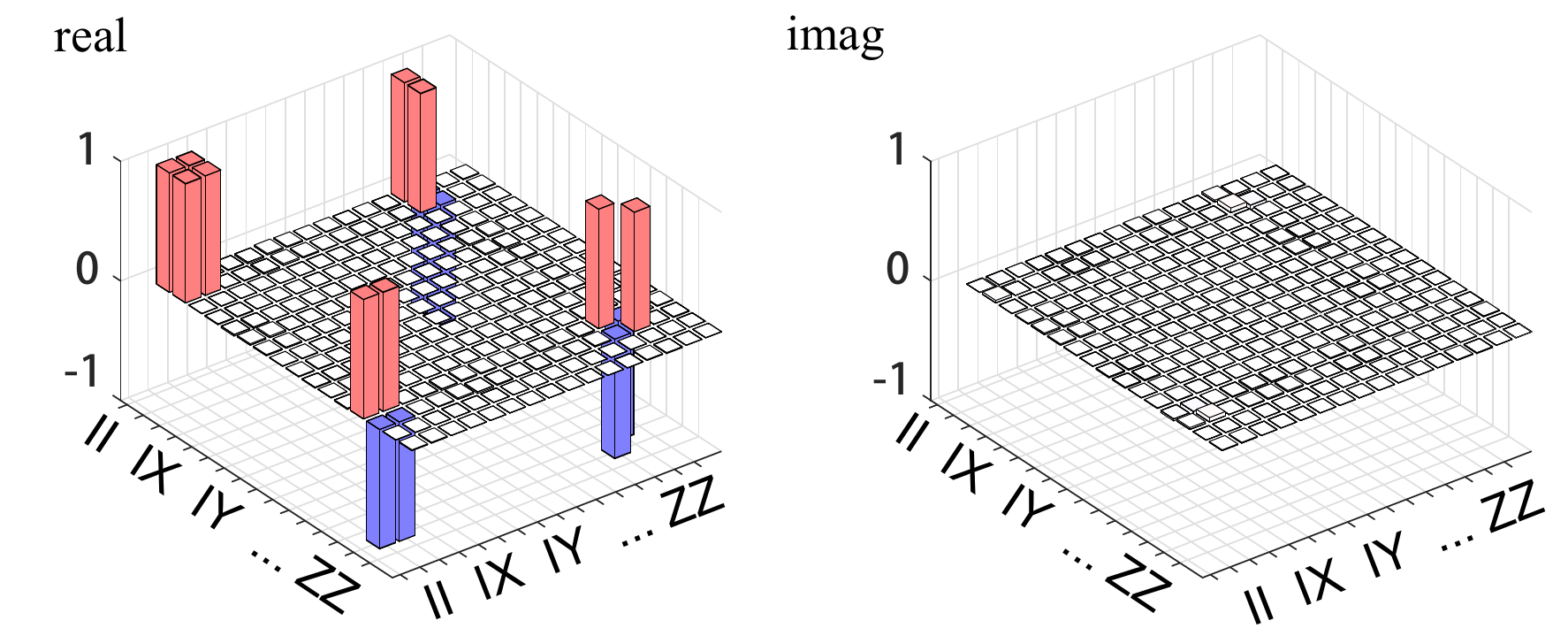}}
	\caption{\label{fig:tomo_cnot}
    The result of QPT of the {\sc cnot} gate.  Here "real" and "imag" denote the real and imaginary parts, respectively.
Bars without color represent
the ideal {\sc cnot} gate, while bars with color represent the actual {\sc cnot} gate. The
average gate fidelity of the {\sc cnot} gate is $F_A = 0.997$.		
    }
\end{figure}
%


%
\section{Process tomography of the {\sc cnot} gate}
To determine the actual average gate fidelity of the {\sc cnot} gate, we perform
QPT on the {\sc cnot} gate. Here  we employ  36 probe
states from the set 
$ \{ |H\rangle, |V\rangle, |D\rangle, |A\rangle, |R\rangle, |L\rangle\}^{\otimes 2}$ and all 9  Pauli measurement settings. 
In each experimental setting (which  means a specific combination of a probe
state and a measurement setting), we perform  20000 
measurements in total (the large number of measurements are employed to suppress the statistical fluctuation). Then the maximum likelihood method is employed to infer the
process matrix of the {\sc cnot} gate from the measurement data \cite{QPT}. The result of QPT is shown in Fig.~\ref{fig:tomo_cnot}, from which we can determine the average gate fidelity
of the {\sc cnot} gate, with the result $0.997$.



%
\bibliography{cites}